\documentclass[aps,prd,preprint,preprintnumbers,showpacs,showkeys,nofootinbib,superscriptaddress]{revtex4-1}

\usepackage{booktabs}    
\usepackage{placeins}
\usepackage{soul}
\usepackage[usenames,dvipsnames,svgnames]{xcolor}
\definecolor{redd}{rgb}{0.8, 0.1,0.2}
\definecolor{navy}{rgb}{0.05, 0.23,0.75}
\usepackage[colorlinks=true,citecolor=red,linkcolor=blue]{hyperref}
\usepackage{chngcntr}
\hypersetup{
     colorlinks   = true,
     citecolor    = navy,
	linkcolor = redd,
	urlcolor=navy,
	anchorcolor=blue
}
\usepackage{bbm}
\usepackage{amsfonts}
\usepackage{amsmath,amssymb}
\usepackage{mathrsfs}
\usepackage{epsfig}
\usepackage{graphicx}
\usepackage{wrapfig}                 
\usepackage{url}
\usepackage{hyperref}
\usepackage{float}
\usepackage{pstricks}
\usepackage{color}
\usepackage{multirow}
\usepackage{lipsum}
\usepackage{enumitem}
\usepackage{ctable}
\newcolumntype{L}{>{\centering\arraybackslash}m{1.5cm}}

\usepackage{enumitem}

\usepackage{chngcntr}

\newcommand{\bear}{\begin{array}}
\newcommand {\eear}{\end{array}}

\newcommand{\beq}{\begin {equation}}
\newcommand{\eeq}{\end   {equation}}
\newcommand{\bea}{\begin {eqnarray}}
\newcommand{\eea}{\end   {eqnarray}}
\newcommand{\baa}{\begin {array}   }
\newcommand{\eaa}{\end   {array}   }
\newcommand{\bit}{\begin {itemize} }
\newcommand{\eit}{\end   {itemize} }
\newcommand{\be }{\begin {equation}}
\newcommand{\ee }{\end   {equation}}
\newcommand{\nn }{\nonumber        }

\def\bea{\begin{eqnarray}}
\def\eea{\end{eqnarray}}

\newcommand{\bef}{\begin{figure}}
\newcommand {\eef}{\end{figure}}
\newcommand{\bec}{\begin{center}}
\newcommand {\eec}{\end{center}}

\definecolor{cerulean}{rgb}{0., 0.62,0.7}

\newcommand{\twiddle}{{\raise.17ex\hbox{$\scriptstyle\sim$}}}

\begin{document}
\begin{titlepage}

\title{EFTs meet Higgs Nonlinearity, Compositeness and (Neutral) Naturalness}

\author{Hao-Lin Li}
\email{lihaolin@itp.ac.cn}
\affiliation{CAS Key Laboratory of Theoretical Physics, Institute of Theoretical Physics, Chinese Academy of Sciences, Beijing 100190, P. R. China}

\author{Ling-Xiao Xu}
\email{lingxiaoxu@pku.edu.cn}
\affiliation{Department of Physics and State Key Laboratory of Nuclear Physics and Technology, Peking University, Beijing 100871, China}

\author{Jiang-Hao Yu}
\email{jhyu@itp.ac.cn}
\affiliation{CAS Key Laboratory of Theoretical Physics, Institute of Theoretical Physics, Chinese Academy of Sciences, Beijing 100190, P. R. China}
\affiliation{School of Physical Sciences, University of Chinese Academy of Sciences, No.19A Yuquan Road, Beijing 100049, P.R. China}

\author{Shou-hua Zhu}
\email{shzhu@pku.edu.cn}
\affiliation{Department of Physics and State Key Laboratory of Nuclear Physics and Technology, Peking University, Beijing 100871, China}
\affiliation{Collaborative Innovation Center of Quantum Matter, Beijing, 100871, China}
\affiliation{Center for High Energy Physics, Peking University, Beijing 100871, China}

\begin{abstract}
Composite Higgs and neutral-naturalness models are popular scenarios in which the Higgs boson is a pseudo Nambu-Goldstone boson (PNGB), and naturalness problem is addressed by composite top partners. Since the standard model effective field theory (SMEFT) with dimension-six operators cannot fully retain the information of Higgs nonlinearity due to its PNGB nature, we systematically construct low energy Lagrangian in which the information of compositeness and Higgs nonlinearity are encoded in the form factors, the two-point functions in the top sector. We classify naturalness conditions in various scenarios, and first present these form factors in composite neutral naturalness models. After extracting out Higgs effective couplings from these form factors and performing the global fit, we find the value of Higgs top coupling could still be larger than the standard model one if the top quark is embedded in the higher dimensional representations. Also we find the impact of Higgs nonlinearity is enhanced by the large mass splitting between composite states. In this case, pattern of the correlation between the $t\bar{t}h$ and $t\bar{t}hh$ couplings is quite different for the linear and nonlinear Higgs descriptions. 
\end{abstract}

\maketitle

\end{titlepage}
\section{Introduction}
After the discovery of the Higgs boson, the lack of the evidence of new physics and the precision measurement of the Higgs properties have already pushed the cut-off scale of the Standard Model (SM) up to TeV if we view it as an effective field theory (EFT), thereby leaving the origin of the smallness of the electroweak (EW) scale and the question whether the ultraviolet (UV) theory is weakly-couple or strongly-coupled as mysteries. To be specific, the nature of the Higgs boson is still unknown. One of the most theoretically-motivated scenarios is to treat the Higgs boson as pseudo Nambu-Goldstone boson (PNGB) emerging from spontaneously broken global symmetry at TeV scale~\cite{Kaplan:1983fs,Kaplan:1983sm,Dugan:1984hq}, or in contrast it can just be a SM-like fundamental scalar. 
For the case of PNGB Higgs, the Higgs boson transforms nonlinearly in the coset space, exhibiting the curvature of this space~\cite{Coleman:1969sm,Callan:1969sn,Alonso:2016btr,Alonso:2016oah}, which is denoted as the Higgs nonlinearity.

\begin{figure}[tb]
\includegraphics[scale=0.5]{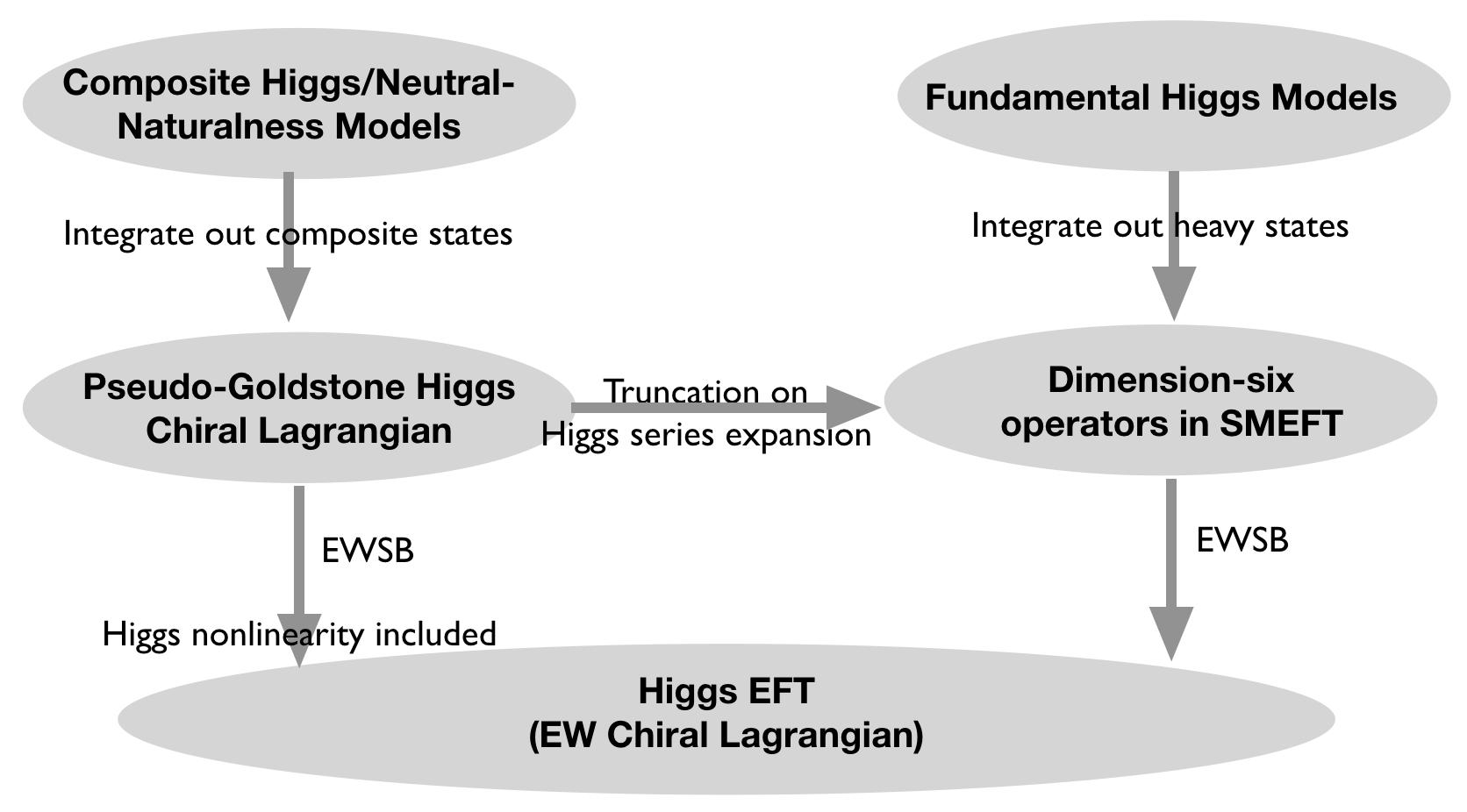}
\caption{Illustration of the relations between different EFTs.} 
\label{fig:frameworks}
\end{figure}

Since there is no significant evidence of new physics observed so far at the Large Hadron Collider (LHC), it is highly motivated to study phenomena involving only the SM particles within the framework of effective theories. In the top-down approach, one can use the techniques, such as equation of motion, or covariant derivative expansion~\cite{Henning:2014wua}, to derive effective theories by directly integrating out the heavy degrees of freedom. One of the most popular EFT frameworks is the SMEFT~\cite{Buchmuller:1985jz,Grzadkowski:2010es,Giudice:2007fh}, which inherits the SM gauge symmetries and parameterizes new physics effects by a cut-off scale $\Lambda$ and Wilson coefficients of high dimensional local operators. For the fundamental Higgs theories, all the heavy particles can be integrated out and thus decoupled, the low energy theory is well approximated by the SMEFT with dimensional-six operators. However, up to dimension-six, the effective operators in SMEFT do not fully capture the information of the Higgs nonlinearity. Thus if the UV theory is strongly coupled and the Higgs is a PNGB, one has to resum operators to all order of ${\mathcal O}(v^2/\Lambda^2)$ to recover the full Higgs nonlinearity effects, which is quite inefficient in the SMEFT. 
A better way is using the CCWZ formalism~\cite{Coleman:1969sm,Callan:1969sn}, which maintains the Higgs nonlinearity effect, to construct the chiral Lagrangian order by order below the mass scale of composite states $m_\rho$, with the chiral expansion ${\mathcal O}(E^2/m_\rho^2)$ if the typical energy transfer $E$ is much smaller than $m_\rho$~\cite{Contino:2011np,Alonso:2014wta,Liu:2018vel,Liu:2018qtb}. In composite Higgs~\cite{Agashe:2004rs} and neutral naturalness~\cite{Chacko:2005pe,Craig:2015pha} models, the UV dynamics is strongly coupled, and contains composite states. After integrating out heavy composite states, one obtains the low energy chiral Lagrangian in which the Higgs boson is parametrized as one of the PNGBs in the coset space. For convenience, this EFT is dubbed as ``PNGB Higgs chiral Lagrangian". Within each order of chiral expansion, only after truncating the series expansion of Higgs field up to a certain order, the high dimensional local operator of SMEFT can be matched on. It is this procedure that renders the nonlinearity of PNGB Higgs somewhat lost in the dimensional-six Lagrangian of the SMEFT. After electroweak symmetry breaking (EWSB), one can expand the Higgs field in both SMEFT and PNGB Chiral Lagrangian around the vacuum expectation value (VEV) and match to the effective Higgs couplings defined in Higgs EFT (HEFT)~\cite{Appelquist:1980vg,Longhitano:1980iz,Feruglio:1992wf,Koulovassilopoulos:1993pw,Grinstein:2007iv,Contino:2010mh,Alonso:2012px,Buchalla:2013rka,Buchalla:2013eza,Buchalla:2014eca}, in which the Higgs boson is a singlet scalar with EWSB and the coset space only includes the longitudinal $W$ and $Z$ bosons. These effective Higgs couplings are directly related to the Higgs coupling measurements at the LHC.  The relation between these EFTs is depicted in Fig.~\ref{fig:frameworks}. Note that by matching PNGB chiral Lagrangian directly on the Higgs couplings in HEFT, the Higgs nonlinearity effect is kept to all orders. 

\begin{figure}[b]
\includegraphics[scale=0.35]{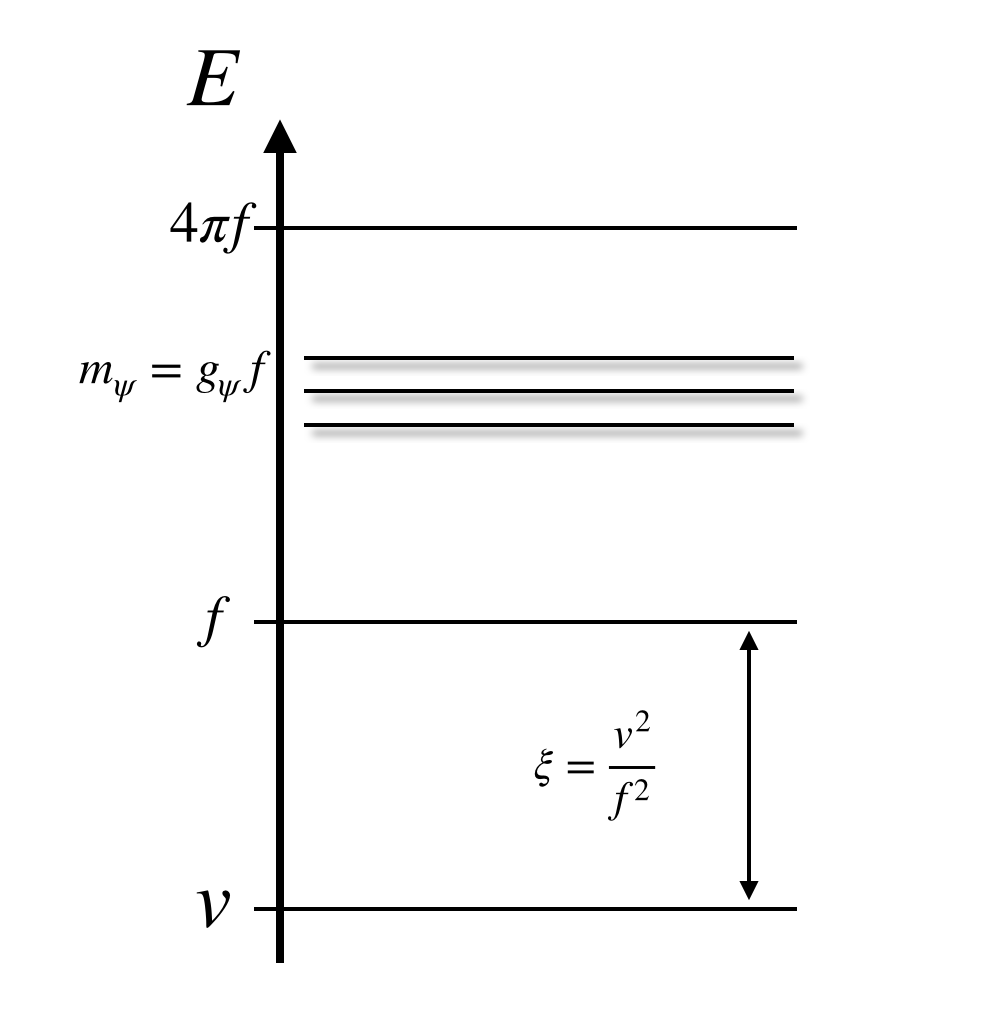}
\includegraphics[scale=0.35]{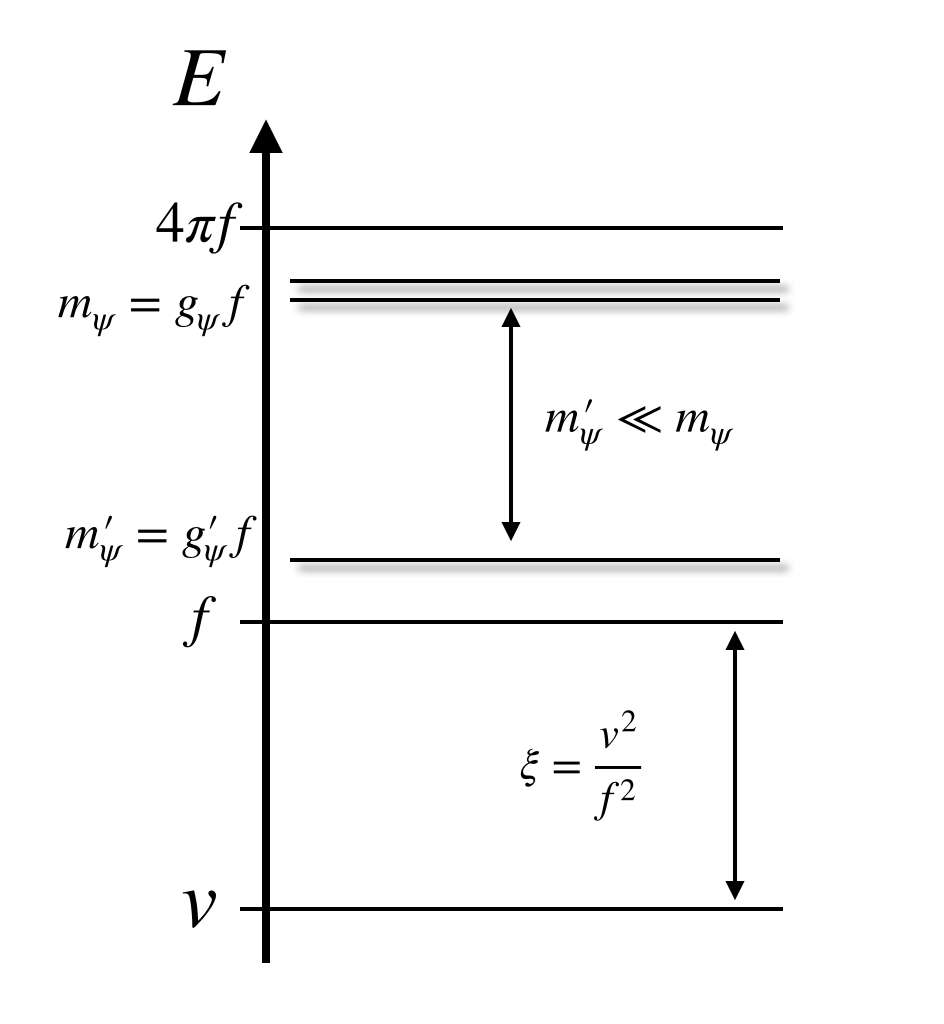}
\caption{\small The mass spectrum for composite states. All the composite states are normally expected at the mass scale $m_\psi$ (left panel), or there are large mass splittings between different composite states (right panel). Compared to the case that all the masses are almost at $m_\psi$, we find the impact of Higgs nonlinearity can be enlarged with the mass spectrum as shown in the right panel.} 
\label{fig:eft}
\end{figure}

In this paper, we aim to systematically study patterns of Higgs effective couplings caused by Higgs nonlinearity and compositeness in the general framework of composite Higgs/neutral-naturalness scenarios. These scenarios are usually constructed under the paradigm of partial compositeness~\cite{Kaplan:1991dc,Contino:2006nn}, namely the Lagrangian consists of three parts: the elementary sector, the composite sector and the mixing sector. To be specific, the model spectrum contains the elementary SM particles and the composite states, and thus we have
\bea
\mathcal{L}_{\textrm{total}} = \mathcal{L}_{\textrm{composite}} + \mathcal{L}_{\textrm{elementary}} + \mathcal{L}_{\textrm{mix}}.
\eea
To see the impact of $\mathcal{L}_{\textrm{mix}}$ on Higgs coupling deviations below the scale of composite resonances, it is convenient to use ``form factors", defined as the two point functions of the elementary fields, to parametrize the information of spectrum of composite particles and the Higgs nonlinearity after integrating out composite states. In contrast to the local operators defined in SMEFT or PNGB Higgs chiral Lagrangian, these two point functions contain non-trivial momentum dependence $Q^2$ from which one can derive the Higgs potential~\cite{Contino:2010rs,Pomarol:2012qf,Marzocca:2012zn}. 
Higgs couplings in HEFT are nevertheless derived by taking the low energy limit as $Q^2\to 0$ without using PNGB Higgs chiral Lagrangian.
The deviations of the Higgs couplings from the SM values exhibit the Higgs nonlinearity effects, characterized by the ratio of the EW scale $v$ and the global symmetry breaking scale $f$. 
Interestingly, we find the impact of Higgs nonlinearity is enlarged in Higgs couplings when composite states have large mass splittings (right panel of Fig.~\ref{fig:eft}), including both the mass splitting between full composite multiplets, and the splitting inside any individual composite multiplet caused by mixing with elementary particles. In contrast, it is normally expected that there is roughly only one mass scale $m_\rho$ for all the composite states (left panel of Fig.~\ref{fig:eft}).

In this work we focus on the low energy Lagrangian and its phenomena in the Higgs sector in various composite Higgs model with and without hidden sectors, with fermions embedded in fundamental and higher dimensional representations. These include minimal composite Higgs models (MCHM)~\cite{Agashe:2004rs,Contino:2006qr}, composite twin Higgs models (CTHM)~\cite{Geller:2014kta,Barbieri:2015lqa,Low:2015nqa} and composite minimal neutral naturalness model (CMNNM)~\cite{Xu:2018ofw}. 
Instead of studying the low energy theories model by model, we organize the low energy Lagrangian in a general way, and several works are in order:
\bit
\item
We organize several naturalness conditions that can be realized in the top sector in a general manner. One of the following symmetries: collective symmetry, left-right parity, and mirror parity, can be imposed to  eliminate quadratic divergence in the top sector. 
\item Then we analyze the PNGB-Higgs dependence of form factors in a universal way without any detailed information from the UV models, which is a generalization of the form factor method in literatures. 
\item We are the first to present expressions of the form factors in the composite twin Higgs and minimal neutral naturalness models. 
\item The Higgs effective couplings in the HEFT are derived systematically using general form factors, in which the information of Higgs nonlinearity effect and the spectrum of composite states is encoded.
\item Finally we perform the global fit on the Higgs couplings with the latest $tth$ data, and obtain numerical results which could serve as a theoretical guidance for the future Higgs coupling measurements.
\eit

The paper is organized as follows. In Sec.~\ref{sec:nat}, we list several naturalness conditions from which several different composite models are motivated. In Sec.~\ref{sec:ch}, we lay out the general framework of the form factors and discuss their general properties from a bottom-up perspective. In Sec.~\ref{sec:coupling}, we derive all the Higgs couplings based on general form factors. In Sec.~\ref{sec:cons}, the experimental constraints are discussed. In Sec.~\ref{sec:num}, we present details on numerical studies and parameter scan. Finally we conclude in Sec.~\ref{sec:con} with all the results of form factors in specific models and other supplemental details being collected in Sec.~\ref{sec:app}.

\section{Natural top quark sector}
\label{sec:nat}

In this work, we focus on the properties of Higgs boson, such as nature of Higgs and Higgs couplings, in the composite Higgs framework. Using the PNGB Higgs chiral Lagrangian, up to the $\mathcal{O}(p^2)$ order, the Higgs couplings to the $W$ and $Z$ bosons are universal, which is not affected by integrating out heavy vector resonances, as presented in the App.~\ref{app:bosonic}. On the other hand, the Higgs couplings to fermions depend on the fermion embedding, and the Higgs potential is radiatively generated by the loop corrections in the fermion sector, especially the top quark sector. Therefore, the fermion embedding is essential to the form of the Higgs couplings in composite Higgs and neutral naturalness models.

Furthermore, there are special requirements on the fermion embedding in the composite Higgs model. In the original composite Higgs model proposed in 1980s~\cite{Kaplan:1983fs,Kaplan:1983sm,Dugan:1984hq}, large fine tuning was required to make the scale separation $f \gg v$, because there is no special symmetry in the fermion sector to cancel the quadratic dependence on $\Lambda$ from the top quark loop. In 2000s, the old idea of PNGB Higgs has been revived~\cite{ArkaniHamed:2001ca,ArkaniHamed:2001nc} due to the collective symmetry breaking imposed in the fermion sector. Same idea was applied to minimal composite Higgs model~\cite{Agashe:2004rs}. So we will focus on the fermion sector in the composite Higgs framework, with naturalness conditions imposed.  After realizing the naturalness condition, the Higgs mass (hence the electroweak scale) is therefore at most logarithmically sensitive to the cutoff scale $\Lambda$.
Because of the large top Yukawa coupling, the top sector contributes the most to the Higgs potential among all the SM fermions. Symmetries can relate the top quark to the so-called top partners in such a way that naturalness is realized, which is dubbed as natural top quark sector.

In the composite Higgs framework, the fermionic sector of composite Higgs models can be systematically constructed under the paradigm of partial compositeness~\cite{Kaplan:1991dc,Contino:2006nn}. In this framework, the SM fermions are regarded as the mixed states of elementary fermions and their composite counterparts. To be specific, we have the following Lagrangian that denotes the mixing between elementary and composite particles as
\bea
\mathcal{L}_{\textnormal{mix}}=y_L \bar{\psi}_L\mathcal{O}_R+y_R \bar{\psi}_R\mathcal{O}_L+\textnormal{h.c.}\ ,
\eea
where $\psi$ is elementary fermions external to the composite sector, while $\mathcal{O}$ is the operator only consisting of composite fields, precisely the PNGB Higgs and composite partners. The couplings $y_{L,R}$ denote the strength of mixing between $\psi_{L,R}$ and $\mathcal{O}_{R,L}$, respectively. The shift symmetry of PNGB Higgs is usually explicitly broken due to the mixings, and hence the non-derivative coupled Yukawa couplings as well as the Higgs potential can be generated from the above Lagrangian. The larger the corresponding Yukawa coupling, the larger the mixing angle between the composite and  elementary sector will be, hence the third generation fermions are the most relevant for our consideration. 

Under the paradigm of partial compositeness, the composite sector contains the PNGB Higgs and the top partners that are responsible for eliminating the quadratic divergence. Although conceptually easy, it is nontrivial to realize the naturalness conditions in concrete models technically. Usually various symmetries are imposed as naturalness conditions.
The general mixing Lagrangian for the top sector is parameterized as
\begin{align}
\mathcal{L}_{\textnormal{mix}}&=y_Lf\bar{t}_L\left(a s_h (T_s)_R+b c_h (T_c)_R\right)+y_R f\bar{t}_R \left(a^\prime c_h (T_s)_L+b^\prime s_h (T_c)_L\right)+\textnormal{h.c.}\nn\\
&\ +\widetilde{y}_Lf\bar{\widetilde{t}}_L\left(\widetilde{a} c_h (\widetilde{T}_s)_R+\widetilde{b} s_h (\widetilde{T}_c)_R\right)+\textnormal{h.c.}\ ,
\label{eq:mix}
\end{align}
where the first line denotes the SM sector while the second line denotes the possible existing hidden sector. Here $T_s, T_c, \widetilde{T}_s, \widetilde{T}_c$ denote the composite fermions with which the elementary fermions $t_L, t_R, \widetilde{t}_L$ are mixed after EWSB. $s_h$ and $c_h$ are the shorthand notations for $s_h\equiv\sin(h/f)=\sin\left(2\sqrt{H^\dagger H}/f\right)$ and $c_h\equiv\cos(h/f)=\cos\left(2\sqrt{H^\dagger H}/f\right)$.
For our purpose, we only include the left-handed part for the hidden sector in the above equation, one can generalize it to include the right-handed part once the embedding of $\widetilde{t}_R$ is specified. Note that the embedding of $\widetilde{t}_R$ is not trivial in concrete models, e.g. CTHM~\cite{Geller:2014kta,Low:2015nqa}. Kinetic terms and mass terms for the composite partners are omitted, as they are irrelevant for realizing the naturalness condition.
By the $SU(2)_L$ doublet nature of $s_h$ and singlet nature of $c_h$, it is not hard to see that $T_s$ and $T_c$ belong to $SU(2)_L$ singlet and doublet respectively. Similarly, composite $\widetilde{T}_s$ and $\widetilde{T}_c$ belong to $SU(2)_L$ singlet and doublet respectively. As we will see, the general Lagrangian in Eq.~\ref{eq:mix} can be realized in concrete MCHM or CTHM depending on whether the hidden sector exists.

\begin{figure}[!h]
\includegraphics[scale=0.4]{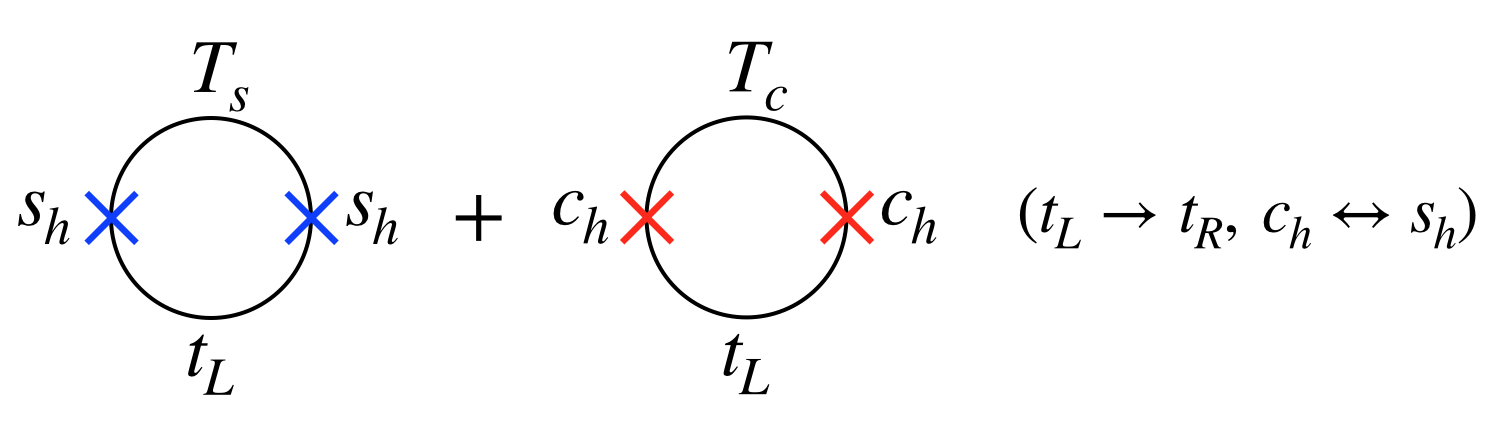}
\caption{Naturalness condition by collective symmetry, which is realized in the two-site composite model. Quadratic divergence is cancelled as $V(h)\sim\Lambda^2\cdot (s_h^2+c_h^2)$.}
\label{2site}
\end{figure}

\begin{figure}[!h]
\includegraphics[scale=0.4]{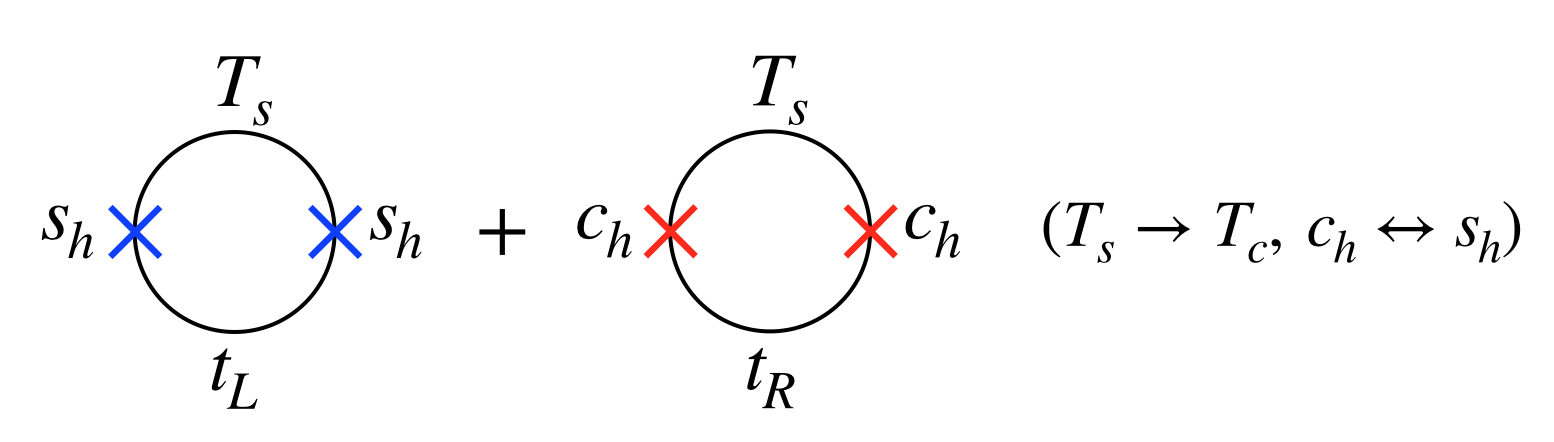}
\caption{Naturalness condition by left-right $Z_2$ symmetry, which is realized if the parity between the left-handed sector and right-handed sector is assigned. Quadratic divergence is cancelled as $V(h)\sim\Lambda^2\cdot (s_h^2+c_h^2)$.}
\label{leftright}
\end{figure}

Let us first focus on the case there is no hidden sector. The mixing Lagrangian can be realized in MCHM based on the coset $SO(5)/SO(4)$~\cite{Agashe:2004rs}. Regrading fermion embeddings, for example, both $q_L=(t_L,b_L)^T$ and $t_R$ can be embedded in the fundamental representation of $SO(5)$, which is then dubbed as $\rm{MCHM}_{5+5}$~\cite{Contino:2006qr}. Other choices are also possible and have been studied in Refs.~\cite{Azatov:2011qy,Gillioz:2012se,Montull:2013mla,Pappadopulo:2013vca,Carena:2014ria,Kanemura:2014kga,Kanemura:2016tan,Liu:2017dsz,Banerjee:2017wmg}, from the perspective of Higgs coupling deviations and Higgs potential. Considering quadratic divergence cancellation, several options are in order:
\bea
\text{collective symmetry}:\ a^2=b^2,\ \ (a^\prime)^2=(b^\prime)^2 ;
\eea
\bea
\text{left-right $Z_2$ symmetry}: \ y_L^2=y_R^2,\ \ a^2=(a^\prime)^2,\ \ b^2=(b^\prime)^2.
\eea
In above equations, we assume all the mixing parameters are real. The Feynman diagrams corresponding to the above two conditions are depicted in Fig.~\ref{2site} and Fig.~\ref{leftright}; they can be realized in the two-site model~\cite{Foadi:2010bu, Panico:2011pw} and the left-right symmetric model, respectively. For the case if the right-handed top quark $t_R$ is fully composite, such as $\rm{MCHM}_{5+1}$, there is no mixing in the right-handed top quark sector, and thus the case of left-right symmetry cannot be realized. The collective symmetry could be realized in the $\rm{MCHM}_{5+1}$ with $a' =0,  b' = 0$.

Let us first illustrate the scenarios with collective symmetry~\cite{Contino:2006qr}. Here we consider the two-site model with the coset $SO(5)_1\times SO(5)_2/SO(5)_V$~\cite{Foadi:2010bu, Panico:2011pw}. 
In the $\rm{MCHM}_{5+5}$ representation, the elementary $t_L$ and $t_R$ can be embedded in the representations of $SO(5)_1$, while composite partners $T_s$ and $T_c$ are embedded within the representations of $SO(5)_2$. They can mix with each other through the link field $\Sigma$ between two sites.  In unitary gauge, the $\Sigma$ field is
\bea
\Sigma=\left(
\baa{ccc}
\bold 1_{3\times 3}&\ &\ \\
\ &c_h&s_h\\
\ &-s_h&c_h\\  
\eaa
\right).
\eea
Then the explicit mixing terms in the two-site model is
\bea
\mathcal{L}_{2\textrm{-site}}=y_L\bar{Q}_L\Sigma\Psi_R+y_R \bar{Q}_R\Sigma\Psi_L+\textnormal{h.c.}\ ,
\eea
where $Q_L$ and $Q_R$ are $5$-plets under $SO(5)_1$ in which $q_L$ and $t_R$ are embedded, while $\Psi$ is a $5$-plet under $SO(5)_2$ in which $T_c$ and $T_s$ are the fourth and fifth component, respectively. 
Since $T_s$ and $T_c$ arise from a single fermionic multiplet, their mixing parameters equal such that the case of collective symmetry is realized. With collective symmetry breaking, the Higgs field $\Sigma$ can be rotated away if the global symmetry $SO(5)_1$ or $SO(5)_2$ is exact. Typically soft terms are needed to prevent the Higgs boson to be an exact Goldstone particle.
On the other hand, in the $\rm{MCHM}_{5+1}$ representation, the right-handed top quark $t_R$ is fully composite and a singlet under $SO(4)_2$~\cite{Csaki:2018zzf}. Thus 
the Lagrangian is written as
\bea
\mathcal{L}_{2\textrm{-site}}=y_L\bar{Q}_L\Sigma\Psi_R+y_Rf\bar{t}_R\Psi_{1L}+\textnormal{h.c.}\ ,
\eea
where there is no Higgs dependence on the $t_R$ term.

The scenario with left-right symmetry has not yet been studied in the literature. The left-right parity is realized if we assume the theory is invariant under the following transformation
\bea
\textnormal{left-right}\ Z_2\ \textrm{symmetry}:\ \ \ t_L\leftrightarrow t_R, \ \ \ (T_{s,c})_L\leftrightarrow (T_{s,c})_R, \ \ \ s_h\leftrightarrow c_h.
\eea
As the above symmetry assignment explicitly relates the left-handed sector with the right-handed sector, it is named as the left-right parity. To be specific, we consider the $SO(6)/SO(5)$ coset and this parity can be realized in the following Lagrangian 
\bea
\mathcal{L}_{\textrm{LR}}&=y_Lf (\bar{q}_L)^i \left[\Sigma_{iJ} \Psi_5^J+\Sigma_{i6}\Psi_1\right]+y_Rf (\bar{q}_R)^i \left[\Sigma_{iJ} \Psi_5^J+\Sigma_{i6}\Psi_1\right],
\eea
where $\Sigma$ is the Goldstone matrix generalized to the $SO(6)/SO(5)$ coset, $T_{s,c}$ are components inside $\Psi_5, \Psi_1$ respectively, and accordingly the $t_L$ and $t_R$ can be embedded into the fundamental representation
\bea
q_L=\frac{1}{\sqrt{2}}\left(ib_L, b_L, it_L, -t_L, 0, 0\right)^T, \ q_R=\frac{1}{\sqrt{2}}\left(0, 0, 0, 0, t_R, t_R\right)^T\ .
\eea
In the $SO(5)/SO(4)$ coset, there is a factor $\sqrt{2}$ in $q_R$ which can cause complications in the normalization of the kinetic term.

\begin{figure}[!h]
\includegraphics[scale=0.4]{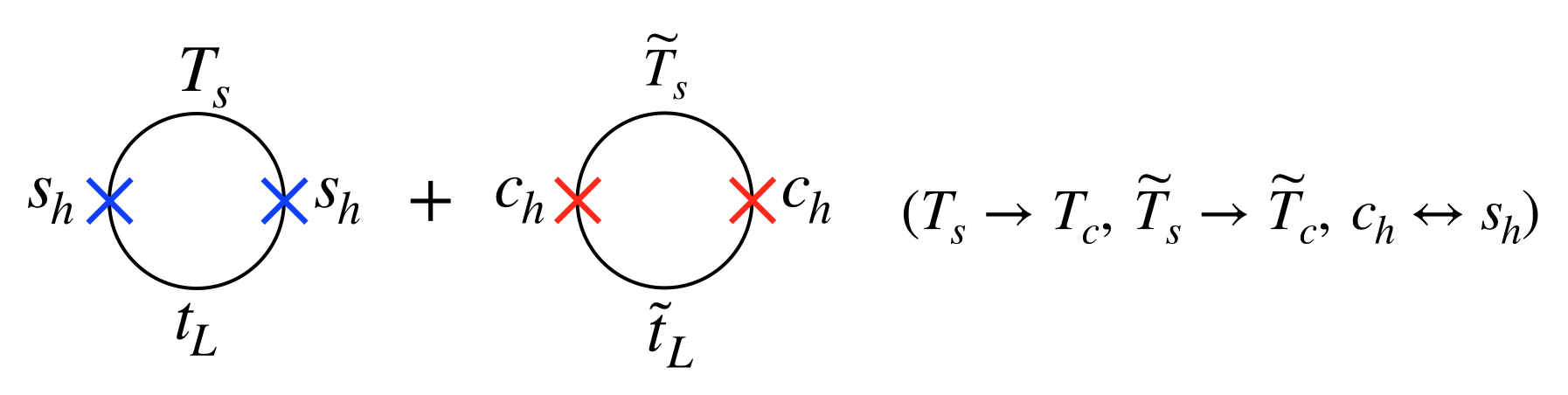}
\caption{Naturalness condition by mirror $Z_2$ symmetry is realized if the mirror parity between the visible sector and hidden sector is assigned. Quadratic divergence is cancelled as $V(h)\sim\Lambda^2\cdot (s_h^2+c_h^2)$.}
\label{twin}
\end{figure}

Hidden sectors can possibly exist in addition to the visible sector (or SM sector), and it can contribute to the Higgs potential. In this case, the naturalness condition yields the relation between couplings:
\bea
\text{mirror $Z_2$ symmetry}:\ y_L^2=\widetilde{y}_L^2,\  a^2=\widetilde{a}^2,\ b^2=\widetilde{b}^2 ,
\eea 
if the parity between the hidden sector and visible sector is assigned. Parameters $a^\prime$ and $b^\prime$ could be assumed to be zero.
This case is depicted in Fig.~\ref{twin}, and it can be realized in the twin Higgs model~\cite{Chacko:2005pe}.
One typically requires the global symmetry groups larger than $SO(5)$ to accommodate the extra hidden fermions.
In this paper, we will systematically study CTHM with the coset $SO(8)/SO(7)$~\cite{Geller:2014kta,Barbieri:2015lqa,Low:2015nqa}. Similar constructions are realized in the coset of $SO(6)/SO(5)$~\cite{Serra:2017poj,Csaki:2017jby} due to the existence of trigonometric parity, and the most minimal coset that can accommodate the trigonometric parity is $SU(3)/SU(2)$~\cite{Csaki:2017jby} if custodial symmetry is not required.
With the presence of the hidden sector, a $Z_2$ mirror parity can be assigned explicitly between the SM sector and the hidden sector (or the mirror sector) as
\bea
\textnormal{mirror}\ Z_2\ \textrm{symmetry}:\ \ \ t_L\leftrightarrow \widetilde{t}_L, \ \ \ T_{s,c}\leftrightarrow\widetilde{T}_{s,c},\ \ \ s_h\leftrightarrow c_h.
\eea
As an explicit example, the above mirror symmetry can be realized in $\textrm{CTHM}_{8+1}$ as
\bea
\mathcal{L}_{8+1}&=y_Lf (\bar{q}^8_L)^i \left[\Sigma_{iJ} \Psi_7^J+\Sigma_{i8}\Psi_1\right]+\textnormal{twin\ sector}(y_L,\bar{q}_L,\Psi\rightarrow \widetilde{y}_L,\bar{\widetilde{q}}_L,\widetilde{\Psi})\ ,
\eea
where $\Sigma$ is the Goldstone matrix generalized to the $SO(8)/SO(7)$ coset, $T_{s,c}$ and $\widetilde{T}_{s,c}$ are components inside $\Psi_7, \Psi_1$ and $\widetilde{\Psi}_7, \widetilde{\Psi}_1$ respectively.

Although there are many methods as shown above that can be utilized to eliminating quadratic divergence, it is still motivated to find novel ways to realize the realistic Higgs potential. The Higgs potential can be generated radiatively, and vacuum misalignment between the electroweak scale and the scale $f$ is naturally realized even with only infrared (IR) fermionic loop contributions. For that, the elementary top partners in the color-neutral sector may carry electroweak quantum numbers, and the vacuum misalignment is connected to the masses of these particles.
In case that a color-neutral sector with more than one elementary top partner is introduced to realize the idea of neutral naturalness, the Lagrangian of Eq.~\ref{eq:mix} can be further generalized to 
\begin{align}
\mathcal{L}^\prime_{\textnormal{mix}}&=y_Lf\bar{t}_L\left(a s_h (T_s)_R+b c_h (T_c)_R\right)+\textnormal{h.c.}\nn\\
&\ +\widetilde{y}_Lf\left[\bar{\widetilde{t}}_L\left(\widetilde{a} c_h (\widetilde{T}_s)_R+\widetilde{b} s_h (\widetilde{T}_c)_R\right)+\bar{\widetilde{t}^\prime}_L\left(\widetilde{a}^\prime s_h (\widetilde{T}_s)_R+\widetilde{b}^\prime c_h (\widetilde{T}_c)_R\right)\right]+\textnormal{h.c.}\ ,
\label{eq:mix2}
\end{align}
where $\widetilde{t}_L$ is a $SU(2)_L$ singlet while $\widetilde{t}_L^\prime$ belongs to a $SU(2)_L$ doublet.
The minimal model can be realized with the coset $SO(5)/SO(4)$~\cite{Xu:2018ofw}, the same coset utilized in the popular minimal composite Higgs model~\cite{Agashe:2004rs} if custodial symmetry is required. Thus the model in Ref.~\cite{Xu:2018ofw} is dubbed as the minimal neutral naturalness model (MNNM). The quadratic divergence is cancelled as
\bea
V(h)\sim\Lambda^2\cdot \left(\frac{1}{2}s_h^2+\frac{1}{2}s_h^2+c_h^2\right).
\label{eq:cmnnm}
\eea
In its composite extension following the paradigm of partial compositeness, Eq.~\ref{eq:cmnnm} is realized as
\bea
\text{Composite MNNM}: y_L^2=\widetilde{y}_L^2,\ a^2=b^2=(\widetilde{a}^\prime)^2=(\widetilde{b}^\prime)^2=\frac{1}{2}(\widetilde{a})^2=\frac{1}{2}(\widetilde{b})^2;
\eea
This can easily be realized by typical fermion embeddings as shown in Ref.~\cite{Xu:2018ofw}. As we see, quadratic divergence is eliminated by cancellation between the SM sector and the color-neutral sector, and also the composite partners of each individual elementary fermion.
Furthermore, such a framework of the composite neutral naturalness model (CMNNM) will lead to novel Higgs dependence in the color-neutral sector after the composite particles are integrated out.

\section{General Framework}
\label{sec:ch}
Below the scale of compositeness, one can calculate all the low energy observables which can directly be tested at the electroweak scale. Those observables include the Higgs potential and all the Higgs couplings, especially the Higgs coupling to the top quark.
At low energies, one can use form factors to encode the information of composite particles and Higgs nonlinearity. Following the spirit of Ref.~\cite{Contino:2010rs}, the effective Lagrangian of the top sector in momentum space is
\bea
\begin{aligned}
\mathcal{L}_{\textnormal{eff}}&=\bar{t}_L p\!\!\!/\Pi_{t_L}(p^2)t_L+\bar{t}_R p\!\!\!/\Pi_{t_R}(p^2)t_R-\left(\bar{t}_L\Pi_{t_Lt_R}(p^2)t_R+\textnormal{h.c.}\right)\\
&+\bar{\widetilde{t}}_L p\!\!\!/\Pi_{\widetilde{t}_L}(p^2)\widetilde{t}_L+\bar{\widetilde{t}}_R p\!\!\!/\Pi_{\widetilde{t}_R}(p^2)\widetilde{t}_R-\left(\bar{\widetilde{t}}_L\Pi_{\widetilde{t}_L\widetilde{t}_R}(p^2)\widetilde{t}_R+\textnormal{h.c.}\right).
\label{effminimal}
\end{aligned}
\eea
The first line of the above equation denotes the ordinary top sector, while the second line denotes the hidden top sector. We include the hidden sector for generality, although it does not have to exist in specific models. All the $\Pi$ functions are the form factors, and different models in principle can result in different specific form factors. In this section, we focus on the general form of form factors based on several general symmetry arguments rather than derive their expressions in specific models. Note that the form factors in the bosonic sector are discussed in the App.~\ref{app:bosonic}, as they are less relevant to our focus in this paper.

Let us first focus on the form factors $\Pi_{t_L}, \Pi_{t_R}$ and $\Pi_{\widetilde{t}_L}, \Pi_{\widetilde{t}_R}$ that are the chirality-preserving ones.
From a bottom-up perspective, $\Pi_{t_L}, \Pi_{t_R}$ defined above can be organized in powers of $s_h^2$ based on the $SU(2)_L$ doublet nature of Higgs (see e.g. Ref.~\cite{Liu:2017dsz} and others),
\bea
\begin{aligned}
\Pi_{t_L}(-Q^2)&=\Pi_{0t_L}(-Q^2)+\Pi_{1t_L}(-Q^2)\ s^2_h+\Pi_{2t_L}(-Q^2)\ s^4_h+\cdots,\ \label{eq:pitL}\\
\Pi_{t_R}(-Q^2)&=\Pi_{0t_R}(-Q^2)+\Pi_{1t_R}(-Q^2)\ s^2_h+\Pi_{2t_R}(-Q^2)\ s^4_h+\cdots, \label{eq:pitR}
\end{aligned}
\eea
and $\Pi_{\widetilde{t}_L}, \Pi_{\widetilde{t}_R}$ are conveniently expanded in powers of $c_h^2$ accordingly,
\bea
\begin{aligned}
\Pi_{\widetilde{t}_L}(-Q^2)&=\Pi_{0\widetilde{t}_L}(-Q^2)+\Pi_{1\widetilde{t}_L}(-Q^2)\ c^2_h+\cdots,\ \label{eq:pitLtilde}\\
\Pi_{\widetilde{t}_R}(-Q^2)&=\Pi_{0\widetilde{t}_R}(-Q^2)+\Pi_{1\widetilde{t}_R}(-Q^2)\ c^2_h+\Pi_{2\widetilde{t}_R}(-Q^2)\ c^4_h+\cdots. \label{eq:pitRtilde} \\
\end{aligned}
\eea
Given a certain fermion representation, only finite number of the form factors after the above expansion exist. The form factors defined above are already enough to analyze the representations that we consider in this paper, and the dots do not represent omission of higher order contributions. Furthermore, the above equations imply that $\Pi_{t_L}, \Pi_{t_R}$, $\Pi_{\widetilde{t}_L}, \Pi_{\widetilde{t}_R}$ are all $SU(2)_L$ singlets. According to the above definition, the loop momentum has already been Wick-rotated to Euclidean space as $Q^2=-p^2$. For MCHM, all the form factors in the hidden sector are fixed to zero as the hidden sector does not exist. For CTHM, on the other hand, the mirror parity in the top sector relates not only the Higgs dependence as $s_h\leftrightarrow c_h$ between two sectors, but also the form factors after expansion. To be specific, the mirror parity would enforce that $\Pi_{0t_L}=\Pi_{0\widetilde{t}_L}, \Pi_{1t_L}=\Pi_{1\widetilde{t}_L}$,  $\Pi_{0t_R}=\Pi_{0\widetilde{t}_R}, \Pi_{1t_R}=\Pi_{1\widetilde{t}_R}, \Pi_{2t_R}=\Pi_{2\widetilde{t}_R}$. Note that $\Pi_{2t_L}$ is included in the visible sector, but not for its counterpart in the hidden sector. This is because $t_L$ (or the left-handed doublet $Q_L=(t_L,b_L)^T$) can be embedded in the symmetric tensor representation ($14$) of $SO(5)$ in MCHM, but it, and accordingly its hidden counterpart $\widetilde{t}_L$, can only be embedded in the fundamental representation ($8$) of $SO(8)$ in CTHM~\cite{Geller:2014kta,Low:2015nqa}. Thus $\Pi_{2\widetilde{t}_L}$ automatically vanishes.

Let us investigate the chirality-flipping form factors next.
For MCHM, depending on specific fermionic embedding in the $SO(5)$ representation, the expansion of $\Pi_{t_Lt_R}$ can nevertheless be different. For example, in the case both left-handed and right-handed top quark are embedded in the fundamental representation of $SO(5)$, $\Pi_{t_Lt_R}$ is expanded as
\bea
\Pi_{t_Lt_R}(-Q^2)&=\Pi_{1t_Lt_R}(-Q^2)\ c_hs_h+\Pi_{2t_Lt_R}(-Q^2)\ c_hs^3_h+\cdots\ .
\eea
Thus $\Pi_{t_Lt_R}$ is a $SU(2)_L$ doublet. 
It turns out that the above expansion of $\Pi_{t_Lt_R}$ is quite general, and it is valid in many cases of top quark embeddings in MCHM, such as $\textnormal{MCHM}_{5+5}$, $\textnormal{MCHM}_{10+10}$ and $\textnormal{MCHM}_{14+14}$. (See App.~\ref{sec:concreteff} for explicit result of the form factors in these models.)
Nevertheless, if the right-handed top is a $SO(5)$ singlet such as in $\textnormal{MCHM}_{5+1}$, $\Pi_{t_Lt_R}$ is expanded as 
\bea
\Pi_{t_Lt_R}(-Q^2)&=\Pi_{1t_Lt_R}(-Q^2)\ s_h+\Pi_{2t_Lt_R}(-Q^2)\ s^3_h+\cdots\ .\label{eq:pitLR}
\eea
Because of the difference between above two expansions, the resulting Higgs coupling deviations in the top sector will be slightly different. Other choices of fermion embeddings in MCHMs are also possible~\cite{Carena:2014ria}.
On the other hand, if the hidden sector exists, the chirality-flipping form factors $\Pi_{t_Lt_R}$ and $\Pi_{\widetilde{t}_L \widetilde{t}_R}$ are
\bea
\begin{aligned}
\Pi_{t_Lt_R}(-Q^2)&=\Pi_{1t_Lt_R}(-Q^2)\ s_h+\Pi_{2t_Lt_R}(-Q^2)\ s^3_h+\cdots\ ,\\
\Pi_{\widetilde{t}_L \widetilde{t}_R}(-Q^2)&=\Pi_{1\widetilde{t}_L\widetilde{t}_R}(-Q^2)\ c_h+\Pi_{2\widetilde{t}_L\widetilde{t}_R}(-Q^2)\ c^3_h+\cdots\ .
\end{aligned}
\label{eq:twin}
\eea
An important argument is in order. Compared to the previous case, there is no ambiguity for the expansion of $\Pi_{t_Lt_R}$ due to different fermion embeddings, namely the Higgs dependence of $c_h$ in $\Pi_{t_Lt_R}$ is forbidden because of the mirror parity. To be specific, mirror particles $\widetilde{t}_{L,R}$ are unambiguisely both $SU(2)_L$ singlets, then Higgs dependence of odd power of $s_h$ (which is known as $SU(2)_L$ doublet) in $\Pi_{\widetilde{t}_L \widetilde{t}_R}$ is not allowed in the mirror sector. In turn, this leads to the fact that $\Pi_{t_Lt_R}$ can be expanded solely in terms of integer powers of $s_h$ because of the mirror parity exchanging $s_h$ with $c_h$ between the two sectors.
We see concrete models such as $\textnormal{CTHM}_{8+1}$, $\textnormal{CTHM}_{8+28}$ and $\textnormal{CTHM}_{8+35}$ satisfy the above form factor expansion (see App.~\ref{sec:concreteff}).
Furthermore, the mirror parity enforces $\Pi_{1t_Lt_R}=\Pi_{1\widetilde{t}_L\widetilde{t}_R},\Pi_{2t_Lt_R}=\Pi_{2\widetilde{t}_L\widetilde{t}_R}$.

Based on Eq.~\ref{effminimal}, Higgs potential can straightforwardly be derived as
\bea
V(h)_{\textnormal{TH}}=-\frac{2N_c}{16\pi^2}\int^{\Lambda^2}_0 \textnormal{d}Q^2 Q^2\left\{\textnormal{log}[\Pi_{t_L}\Pi_{t_R}\cdot Q^2+\Pi^2_{t_Lt_R}]+\textnormal{log}[\Pi_{\widetilde{t}_L}\Pi_{\widetilde{t}_R}\cdot Q^2+\Pi^2_{\widetilde{t}_L\widetilde{t}_R}]\right\}.
\eea
At the low energy limit of $Q^2\rightarrow 0$, masses of the top quark and its twin partner are roughly
\bea
\begin{aligned}
m_t&=\frac{\Pi_{t_Lt_R}(0)}{\sqrt{\Pi_{t_L}(0)\Pi_{t_R}(0)}}\simeq\frac{\Pi_{1t_Lt_R}(0)\ s_h}{\sqrt{\Pi_{0t_L}(0)\ \Pi_{0t_R}(0)}}\ ,\\
m_{\widetilde{t}}&=\frac{\Pi_{\widetilde{t}_L\widetilde{t}_R}(0)}{\sqrt{\Pi_{\widetilde{t}_L}(0)\Pi_{\widetilde{t}_R}(0)}}\simeq\frac{\Pi_{1\widetilde{t}_L\widetilde{t}_R}(0)\ c_h}{\sqrt{\Pi_{0\widetilde{t}_L}(0)\ \Pi_{0\widetilde{t}_R}(0)}}\ .
\end{aligned}
\label{topmass}
\eea
The second equality in the above equation holds if only the leading terms of the expansion are included. Then the ratio of the masses of the top quark and its twin partner is approximately
\bea
\frac{m_t}{m_{\widetilde{t}}}\simeq \frac{\langle s_h\rangle}{\langle c_h\rangle}.
\eea
As we will see later, $\langle s_h\rangle$ (and hence $\langle c_h\rangle$) is directly related to the ratio of the electroweak scale $v$ and the $f$ scale, i.e., $\langle s_h\rangle =v/f\simeq 1/3$.
Given the SM top mass $m_t\sim y_t v/\sqrt{2}$, the mass of top twin partner will be $m_{\widetilde{t}}\sim y_t f/\sqrt{2}$, whose numerical value can be around TeV.

In the case that the color-neutral sector contains several elementary top partners with and without carrying electroweak quantum numbers, novel Higgs dependence in the color-neutral sector can result from the $SU(2)_L$ quantum numbers of these top partners. In this work, we will limit our discussion within the example raised in Ref.~\cite{Xu:2018ofw} with its generalization left to future study. If the color-neutral top sector has one $SU(2)_L$ doublet and one $SU(2)_L$ singlet, the effective Lagrangian is 
\bea
\mathcal{L}_{\text{eff}}=\ &\bar{t}_L p\!\!\!/\Pi_{t_L}t_L+\bar{t}_R p\!\!\!/\Pi_{t_R}t_R-\bar{t}_L\Pi_{t_Lt_R}t_R+\bar{\widetilde{L}} p\!\!\!/\widetilde{\Pi}_{L}\widetilde{L}+\bar{\widetilde{R}} p\!\!\!/\widetilde{\Pi}_{R}\widetilde{R}-\bar{\widetilde{L}}\widetilde{\Pi}_{LR}\widetilde{R}+\textnormal{h.c.}\ ,
\label{holo}
\eea
where $\widetilde{L}\equiv (\widetilde{t}_L,\widetilde{T}_L)^T$ and $\widetilde{R}\equiv (\widetilde{t}_R,\widetilde{T}_R)^T$, with $\widetilde{t}_{L,R}$ arising from the doublet while $\widetilde{T}_{L,R}$ arising from the singlet. Depending on fermion embeddings, the form factors of the SM sector $\Pi_{t_L},\Pi_{t_R},\Pi_{t_Lt_R}$ have the same patterns of Higgs dependence as in MCHM. For example, we assume $\Pi_{t_L},\Pi_{t_R}$ have no Higgs dependence while 
\bea
\Pi_{t_Lt_R}=\Pi_{1t_Lt_R}s_h
\eea
as in the composite minimal neutral naturalness model (CMNNM)~\cite{Xu:2018ofw}. This assumption is explicitly realized if mass splitting between different components of the full composite multiplet is turned off, and $t_R$ is a $SO(5)$ singlet. On the other hand, the form factors of the color-neutral sector $\widetilde{\Pi}_{L},\widetilde{\Pi}_{R},\widetilde{\Pi}_{LR}$ have both the Higgs dependence of $s_h$ and $c_h\simeq 1-s_h^2/2$. For example, the diagonal terms are expanded as
\bea
\widetilde{\Pi}^{ii}=\widetilde{\Pi}_0^{ii}+\widetilde{\Pi}_1^{ii} s_h^2+\cdots\quad\quad (i=1,2)\ ,
\eea
while the off-diagonal terms are 
\bea
\widetilde{\Pi}^{ij}=s_h\left(\widetilde{\Pi}_1^{ij}+\cdots\right) \quad\quad\quad (i,j=1,2\ \text{and}\ i\neq j)\ ,
\eea
which denotes the mixings between the doublet and singlet in the color-neutral sector. In above equations, the index $L, R, LR$ of the form factors is neglected for convenience.

\section{Effective Higgs Couplings}
\label{sec:coupling}

One can define the effective Higgs coupling after EWSB. To be specific, we have the following couplings defined in the Higgs EFT~\cite{Appelquist:1980vg,Longhitano:1980iz,Feruglio:1992wf,Koulovassilopoulos:1993pw,Grinstein:2007iv,Contino:2010mh,Alonso:2012px}:
\bea
\begin{aligned}
\mathcal{L}_{H}=&\frac{\alpha_s}{12\pi}G^a_{\mu\nu}G^{a\mu\nu}\left(c_g \frac{h}{v}+\frac{1}{2} c_{gghh} \frac{h^2}{v^2}+\cdots\right)+\frac{\alpha}{8\pi}F_{\mu\nu}F^{\mu\nu}\left(c_\gamma\frac{h}{v}+\cdots\right)\\
&-\frac{m_t}{v}c_t\bar{t}th-\frac{m_t}{v^2}c_{t\bar{t}hh}\bar{t}th^2-\frac{m_h^2}{2v}c_{3h}h^3-\frac{m_h^2}{8v^2}c_{4h}h^4+\cdots\\
&+\frac{v^2}{4}\text{Tr}\left[\left(D_\mu U\right)^\dagger \left(D^\mu U\right)\right]\left(1+2c_W\frac{h}{v}+\cdots\right),
\end{aligned}
\label{LHiggs2}
\eea
where the SM limit with the fundamental Higgs boson corresponds to the case that $c_t=c_{3h}=c_{4h}=c_g=c_{gghh}=c_\gamma=c_W=1$ while $c_{t\bar{t}hh}=0$. 
Here $\alpha_s=g_s^2/(4\pi)$ and $\alpha=e^2/(4\pi)$, where $g_s$ and $e$ are couplings for QCD interaction and electromagnetic interaction respectively. 
Being different from previous discussion, it is worth noting that $h$ denotes the physical Higgs boson (without VEV) in the above equation.

In the rest part of this section, we derive all the Higgs effective couplings listed in Eq.~\ref{LHiggs2} based on the general framework of composite Higgs discussed in Sec.~\ref{sec:ch}.

\subsection{Higgs Self Couplings}
\label{subsec:pot}
Despite of the global symmetry breaking pattern, the general Higgs potential of the PNGB Higgs can be parametrized by
\bea
V(h)=-\gamma_f s_h^2+\beta_f s_h^4+\cdots\ ,
\eea
with the so-called ``vacuum misalignment"~\cite{Kaplan:1983fs} parameter explicitly defined as
\bea
\xi=\frac{v^2}{f^2}=\sin^2\left(\frac{\langle h\rangle}{f}\right),
\eea
where $v$ is the usual electroweak scale which gives the correct $W^\pm$ and $Z$ mass, $\gamma_f$ and $\beta_f$ are the coefficients determined by the dynamics that is responsible for generating the Higgs potential. 
The condition for EWSB ($\partial V(\langle h\rangle)/\partial \langle h\rangle=0$) and the physical Higgs mass are respectively
\bea
\xi=\langle s^2_h\rangle=\frac{\gamma_f}{2\beta_f}, 
\eea
\bea
m_h^2=\frac{\partial^2V(\langle h\rangle)}{\partial \langle h\rangle^2}=\frac{8\beta_f}{f^2}\xi (1-\xi).
\eea 
With above results, one can see $\gamma_f$ and $\beta_f$ can be re-parameterized by $\xi$ and $m_h$.
More importantly, the Higgs self interactions are
\bea
c_{3h}=\frac{-\frac{1}{6}\frac{\partial^3 V(\langle h\rangle)}{\partial \langle h\rangle^3}}{-\frac{m_h^2}{2v}}=1-\frac{3}{2}\xi+\mathcal{O}(\xi^2),
\eea
\bea
c_{4h}=\frac{-\frac{1}{24}\frac{\partial^4 V(\langle h\rangle)}{\partial \langle h\rangle^4}}{-\frac{m_h^2}{8v^2}}=1-\frac{25}{3}\xi+\mathcal{O}(\xi^2).
\eea 
The ratio of the Higgs self couplings with their SM values $c_{3h}$ and $c_{4h}$ only depend on $\xi$, rather than the coefficients $\gamma_f$ and $\beta_f$ which parametrize the origin of the Higgs potential. Although it is experimentally challenging, measuring $c_{3h}$ and $c_{4h}$ can directly probe the Higgs boson nature.

For minimal composite Higgs of $SO(5)/SO(4)$, EWSB is not automatically guaranteed and it requires $\gamma_f>0$ to trigger EWSB. It has been pointed out that $\gamma_f>0$ is correlated to the sign of $c_t-c_g$~\cite{Liu:2017dsz}. 
However, EWSB automatically happens in composite twin Higgs of $SO(8)/SO(7)$ due to the property of the mirror parity transformation: $s_h\leftrightarrow c_h$.
To be specific, the Higgs potential for composite twin Higgs can be rewritten as
\bea
V(h)_{\textnormal{TH}}=\frac{\beta_f}{2} (c^4_h+s^4_h)=-\beta_f s^2_hc^2_h=-\beta_f s^2_h+\beta_f s^4_h.
\eea
We see the above Higgs potential is invariant under mirror transformation. More importantly, the minus sign necessary to trigger EWSB is automatically generated with $\xi=1/2$. Extra $Z_2$ breaking effects are needed in twin Higgs models for realizing realistic EWSB with $\xi\ll 1$. Following this direction, a recent work~\cite{Xu:2018ofw} shows the construction that naturally realize realistic EWSB with small $\xi$.

\subsection{Higgs Couplings in the Top Sector}
Before deriving the Higgs-top effective couplings in different classes of composite models, it is useful to have some general discussions on the Higgs contact interactions with gluons and top quark.
The contact interaction between the Higgs boson and the gluons $h^{(n)}gg$ can be derived from~\cite{Ellis:1975ap,Shifman:1979eb,Kniehl:1995tn}
\bea
\mathcal{L}^{(g)}_{eff}=\frac{\alpha_s}{24\pi} G^a_{\mu\nu}G^{a\mu\nu}\sum_i\textnormal{log}\ m^2_i(h)
\label{eq:hgg}
\eea
where $m^2_i(h)$ denotes the general Higgs-dependent masses for the fermions circulating in the gluon loop. 
In the SM, the Higgs-dependent top mass is $m_t(h)=y_t (h+v)/\sqrt{2}$ with $y_t=1$ the top Yukawa coupling. Therefore, the contact interaction of $h^{(n)}gg$ induced by the SM top loop is obtained as
\bea
\mathcal{L}^{(g)}_{top}=\frac{\alpha_s}{12\pi} G^a_{\mu\nu}G^{a\mu\nu} \ \textnormal{log}\left(\frac{h+v}{v}\right),
\eea 
after $m_t(h)$ is normalized with the EW scale $v$.
For composite models, the particles circulating in the gluon loop are the top quark and the top partners.
In general, Eq.~\ref{eq:hgg} is expanded as
\bea
\mathcal{L}^{(g)}_{eff}\equiv \frac{\alpha_s}{12\pi} G^a_{\mu\nu}G^{a\mu\nu} \left(c_g \frac{h}{v}+\frac{1}{2} c_{gghh} \frac{h^2}{v^2}+\cdots\right)
\eea
where $c_g$ and $c_{gghh}$ can be derived
\bea
\begin{aligned}
c_g&=v\ \frac{\partial}{\partial \langle h\rangle}\left[\frac{1}{2}\sum_i\textnormal{log}\ m^2_i(h)\right],\\
c_{gghh}&=-v^2\ \frac{\partial^2}{\partial \langle h\rangle^2}\left[\frac{1}{2}\sum_i\textnormal{log}\ m^2_i(h)\right],
\end{aligned}
\eea 
Thus one obtains the contributions of SM top loop to $hgg$ and $hhgg$ couplings are $\alpha_s/(12\pi v)$ and $-\alpha_s/(24\pi v^2)$, respectively.
For composite Higgs models considered in the paper, $\sum_i\textnormal{log}\ m^2_i(h)$ can be generalized by the expression with the general mass matrix of the top sector $\textnormal{log} \left[ \textnormal{Det}M_t^\dagger M_t\right]$, as there are in general off-diagonal entries denoting the mixings between the top and top partners.

The Higgs coupling with the top quark $c_t$ and $c_{t\bar{t}hh}$ can be straightforwardly derived from the Higgs-dependent mass of the top quark,
\bea
c_t=v\frac{\partial}{\partial\langle h\rangle}\textnormal{log}(m_t)\ ,\quad\quad\quad
c_{\bar{t}thh}=\frac{v^2}{2}\frac{1}{m_t} \frac{\partial^2 m_t}{\partial \langle h\rangle^2}\ ,
\eea
where $m_t$ is explicitly derived in Eq.~\ref{topmass}.

In the following, Higgs couplings are derived in terms of form factors explicitly in different classes of models, and we will see the Higgs couplings in the top sector are sensitive to both the Higgs nonlinearity and the heavy resonances.

\subsubsection{Higgs Couplings in Minimal Composite Higgs Models}
In the minimal composite Higgs, we only study models $\textnormal{MCHM}_{5+5}$, $\textnormal{MCHM}_{10+10}$ and $\textnormal{MCHM}_{14+14}$ here, of which the expansion of $\Pi_{t_Lt_R}\sim\Pi_{1t_Lt_R}\ c_hs_h+\cdots$ is valid.
The expansion $\Pi_{t_Lt_R}\sim\Pi_{1t_Lt_R}\ s_h+\cdots$ and the corresponding models, such as $\textnormal{MCHM}_{5+1}$, are more similar to the case of CTHM, which are left to the discussion in the next subsection.

With the leading approximation of $\xi$, the relevant effective Higgs couplings are
\bea
c_t=v\frac{\partial}{\partial\langle h\rangle}\textnormal{log}(m_t)=1-\frac{3}{2}\xi-\xi\left(\frac{\Pi_{1t_L}(0)}{\Pi_{0t_L}(0)}+\frac{\Pi_{1t_R}(0)}{\Pi_{0t_R}(0)}\right)+2\xi \frac{\Pi_{2t_Lt_R}}{\Pi_{1t_Lt_R}}+\mathcal{O}(\xi^2),
\label{eq:ctmchm}
\eea
\bea
c_g=\frac{v}{2}\frac{\partial}{\partial\langle h\rangle}\textnormal{log}\ \textnormal{Det}(M_t^\dagger M_t)=1-\frac{3}{2}\xi+2\xi \frac{\Pi_{2t_Lt_R}}{\Pi_{1t_Lt_R}}+\mathcal{O}(\xi^2).
\eea
In the framework of partial compositeness, it is proved that $\textnormal{Det}(M_t)\propto\Pi_{t_Lt_R}$ up to an overall Higgs-independent factor~\cite{Azatov:2011qy}. This factor is cancelled out when evaluating $c_g$.
Based on the above expressions of $c_t$ and $c_g$, a few comments are in order. First, for models where the form factor $\Pi_{2t_Lt_R}$ vanishes, $c_g$ is insensitive to the information of heavy resonances. Thus measuring $c_g$ is useful for probing Higgs nonlinearity. Second, the sign of $c_t-c_g$ is correlated with the positiveness of $\gamma_f$, regardless of the presence of $\Pi_{2t_Lt_R}$. As EWSB requires $\gamma_f>0$, $c_t-c_g$ is preferred to be negative~\cite{Liu:2017dsz}. Third, with the presence of $\Pi_{2t_Lt_R}$, both $c_t-1$ and $c_g-1$ can be positive, negative and zero. Otherwise, $c_t$ and $c_g$ must be smaller than one when $\Pi_{2t_Lt_R}$ vanishes.

Beyond single Higgs vertices, $c_{\bar{t}thh}$ and $c_{gghh}$ can also be derived following the same method. The results are
\bea
c_{\bar{t}thh}=\frac{v^2}{2}\frac{1}{m_t} \frac{\partial^2 m_t}{\partial \langle h\rangle^2}=-2\xi-\frac{3}{2}\xi \left(\frac{\Pi_{1t_L}(0)}{\Pi_{0t_L}(0)}+\frac{\Pi_{1t_R}(0)}{\Pi_{0t_R}(0)}\right)+3\xi\frac{\Pi_{2t_Lt_R}}{\Pi_{1t_Lt_R}}+\mathcal{O}(\xi^2),
\label{eq:ctthhmchm}
\eea
\bea
c_{gghh}=-\frac{v^2}{2}\frac{\partial^2}{\partial\langle h\rangle^2}\textnormal{log}\ \textnormal{Det}(M_t^\dagger M_t)=1+\xi\left(1-2\frac{\Pi_{2t_Lt_R}}{\Pi_{1t_Lt_R}}\right)+\mathcal{O}(\xi^{3/2}). 
\eea
Both $c_{\bar{t}thh}$ and $c_{gghh}$ are important to the double Higgs production $gg\to hh$. 

Based on the above results, we see that there are strong correlations between different Higgs couplings. For example, considering all the effective couplings we have
\bea
c_{\bar{t}thh}=-\frac{1}{6}c_g+\frac{3}{2}c_t-\frac{1}{6}c_{gghh}-\frac{7+\xi}{6}\ ,
\eea
where the SM limit corresponds to $c_g=c_t=c_{gghh}=1$ and $c_{\bar{t}thh}=0$. In turn, $\xi$ can be re-parametrized by the couplings $c_{3h}$ and $c_{4h}$. Furthermore, considering the correlation between $c_t$ and $c_{t\bar{t}hh}$, we obtain the relation
\bea
\frac{3}{2}c_t-c_{t\bar{t}hh}-\frac{3}{2}+\frac{\xi}{4}=0.\label{eq:ctthhctmchm}
\eea

\subsubsection{Higgs Couplings in Composite Twin Higgs Models}
Analog to minimal composite Higgs models, we then derive all the effective Higgs couplings in composite twin Higgs models of $SO(8)/SO(7)$. With the leading approximation of $\xi$, we obtain
\bea
c_t=1-\frac{\xi}{2}-\xi\ \left(\frac{\Pi_{1t_L}(0)}{\Pi_{0t_L}(0)}+\frac{\Pi_{1t_R}(0)}{\Pi_{0t_R}(0)}\right)+2\xi\frac{\Pi_{2t_Lt_R}}{\Pi_{1t_Lt_R}}+\mathcal{O}(\xi^2),
\label{eq:ctcthm}
\eea
\bea
c_g=1-\frac{\xi}{2}+2\xi\frac{\Pi_{2t_Lt_R}}{\Pi_{1t_Lt_R}}+\mathcal{O}(\xi^2),
\eea
\bea
c_{\bar{t}thh}=-\frac{\xi}{2}-\frac{3}{2}\xi \left(\frac{\Pi_{1t_L}(0)}{\Pi_{0t_L}(0)}+\frac{\Pi_{1t_R}(0)}{\Pi_{0t_R}(0)}\right)+3\xi \frac{\Pi_{2t_Lt_R}}{\Pi_{1t_Lt_R}}+\mathcal{O}(\xi^2),
\eea
\bea
c_{gghh}=1-2\xi\frac{\Pi_{2t_Lt_R}}{\Pi_{1t_Lt_R}}+\mathcal{O}(\xi^{3/2}).
\eea
Comments are in order, including similarities and differences compared to MCHM. First, $c_g$ is only sensitive to Higgs nonlinearity when the form factor $\Pi_{2t_Lt_R}$ vanishes. This is similar to the previous case of MCHM. Second, contrary to MCHM, $c_t-c_g$ and $c_t-1$ can in principle be positive, negative or zero, as the form factor
$\left(\frac{\Pi_{1t_L}(0)}{\Pi_{0t_L}(0)}+\frac{\Pi_{1t_R}(0)}{\Pi_{0t_R}(0)}\right)$ is not constrained by the condition of EWSB.
Fourth, we see the correlation between different Higgs couplings still exists, such as
\bea
c_{\bar{t}thh}=-\frac{1}{2}c_g+\frac{3}{2}c_t-\frac{1}{2}-\frac{1}{2}c_{gghh}\ ,
\eea
and
\bea
\frac{3}{2}c_t-c_{t\bar{t}hh}-\frac{3}{2}+\frac{\xi}{4}=0.\label{eq:ctthhctcthm}
\eea

\subsubsection{Higgs Couplings in Composite Minimal Neutral Naturalness Model}
Based on the assumption that the SM form factors $\Pi_{t_L},\Pi_{t_R}$ have no Higgs dependence while $\Pi_{t_Lt_R}=\Pi_{1t_Lt_R}s_h$, one can derive the Higgs couplings with the top quark. With the leading approximation of $\xi$, we obtain
\bea
c_t=c_g=1-\frac{\xi}{2}+\mathcal{O}(\xi^2),
\label{eq:ctcnnm}
\eea
\bea
c_{\bar{t}thh}=-\frac{\xi}{2}+\mathcal{O}(\xi^2),
\eea
\bea
c_{gghh}=1+\mathcal{O}(\xi^{3/2}).
\eea
One can see it is similar to CTHM when the combinations of form factors vanish. That make senses since different components inside a full composite multiplet is assumed to be completely degenerate.

\subsection{Higgs Couplings with Photons and $W^\pm,Z$}
Similar to the Higgs couplings to gluons, the contact interactions with photons $h^{(n)}\gamma\gamma$ is derived from
\bea
\mathcal{L}^A_{eff}=\frac{\alpha}{4\pi}F_{\mu\nu}F^{\mu\nu}\left(\sum_iQ_i^2\ \textnormal{log}\ m^2_i(h)-\frac{7}{4}\ \textnormal{log}\ m^2_W(h)\right),
\eea
considering both the fermionic and bosonic contributions where $m^2_i(h)$ denotes the Higgs-dependent masses of the top quark and top partners circulating in the photon loop with corresponding electric charge $Q_i$, and $m_W^2(h)$ is the Higgs-dependent mass for $W^\pm$ such that $m_W^2(h)=\frac{g^2}{4}v^2=\frac{g^2f^2}{4} s_h^2$, as shown in appendix~\ref{app:bosonic}.
After expanding $\mathcal{L}^A_{eff}$ as
\bea
\mathcal{L}^A_{eff}=\frac{\alpha}{4\pi}F_{\mu\nu}F^{\mu\nu}\left(c_\gamma\frac{h}{v}+\frac{1}{2}c_{\gamma\gamma}\frac{h^2}{v^2}+\cdots\right)\ ,
\eea
the effective coupling $c_\gamma$ is directly obtained
\bea
c_\gamma\simeq\frac{4 Q_t^2 c_g-J_\gamma\left(\frac{4m_W^2}{m_h^2}\right)c_W}{4 Q_t^2-J_\gamma\left(\frac{4m_W^2}{m_h^2}\right)}.
\eea
In the above equation, we assume that all the top partners have the same electric charge as the top quark for composite models. Here $c_W$ is explicitly
\bea
c_W=\sqrt{1-\xi}
\label{eq:cw}
\eea
and the loop function is
\bea
J_\gamma(x)=2+3x[1+(2-x)f(x)],\ \ \ f(x)=\textnormal{arcsin}^2(x^{-1/2}),
\eea
which would be $J_\gamma(\infty)=7$ at the limit of large $x$. Note that the result of $c_W$ derived from the form factors is consistent with the result derived from the chiral Lagrangian at the order of $\mathcal{O}(p^2)$. Integrating out the composite $\rho$ meson will not contribute to the $\mathcal{O}(p^2)$ operator~\cite{Contino:2011np}. However, integrating out heavy particles that explicitly break the shift symmetry of PNGB Higgs can also cause $c_W$ deviate from the SM value. Fully resolving this effect in $c_W$ from the effect caused by Higgs nonlinearity requires novel method~\cite{Cao:2018cms}. In this paper, we will not consider this more complicated situation.

\section{Experimental Constraints on Higgs Couplings}
\label{sec:cons} 
In this section, we will discuss the sets of experimental data that we use to derive the constraints on the Higgs couplings and parameters in the model classes we study above.

\begin{figure}[htp]
{\includegraphics[width=0.40\textwidth]{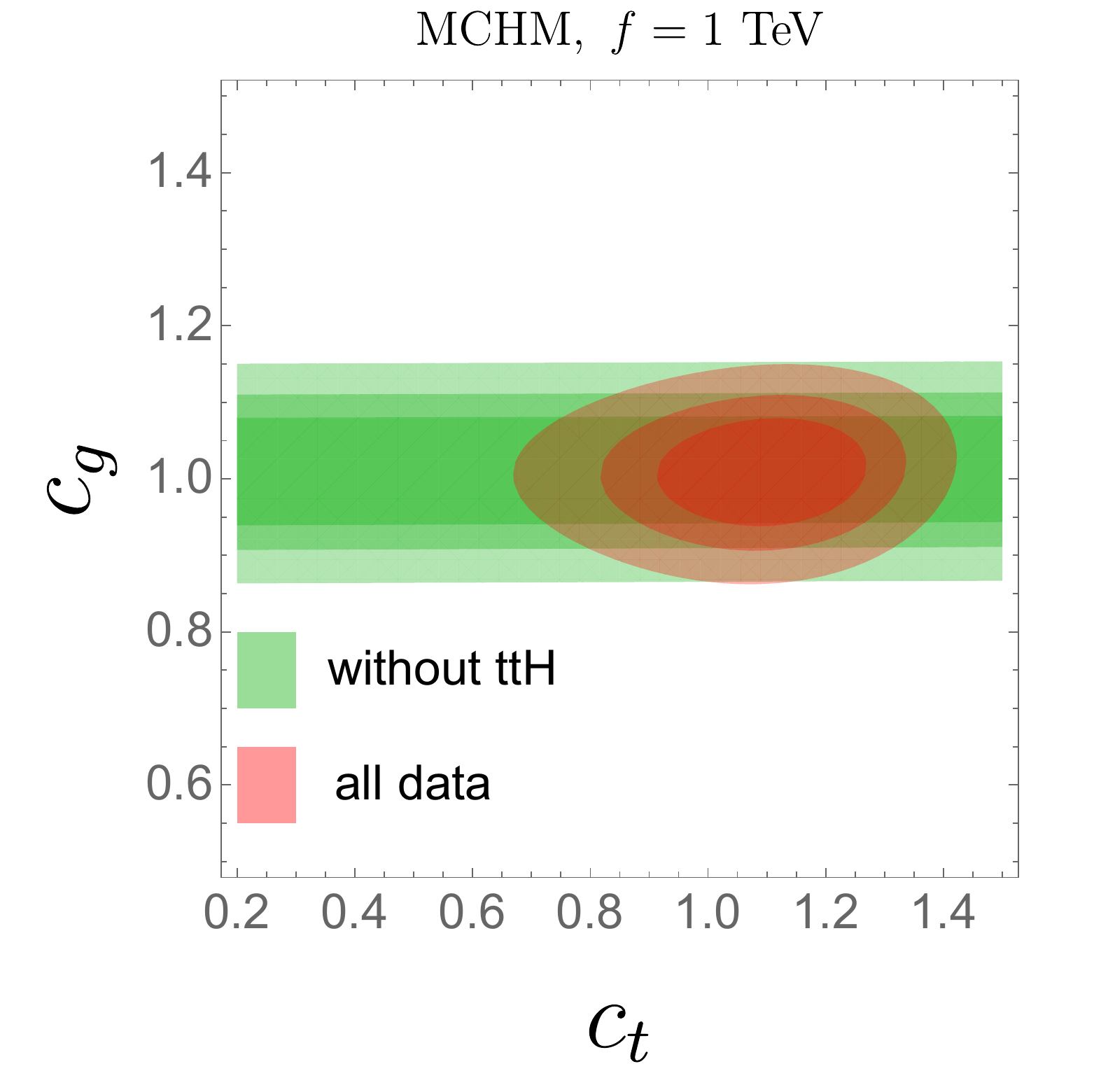}}%
 \quad
{\includegraphics[width=0.40\textwidth]{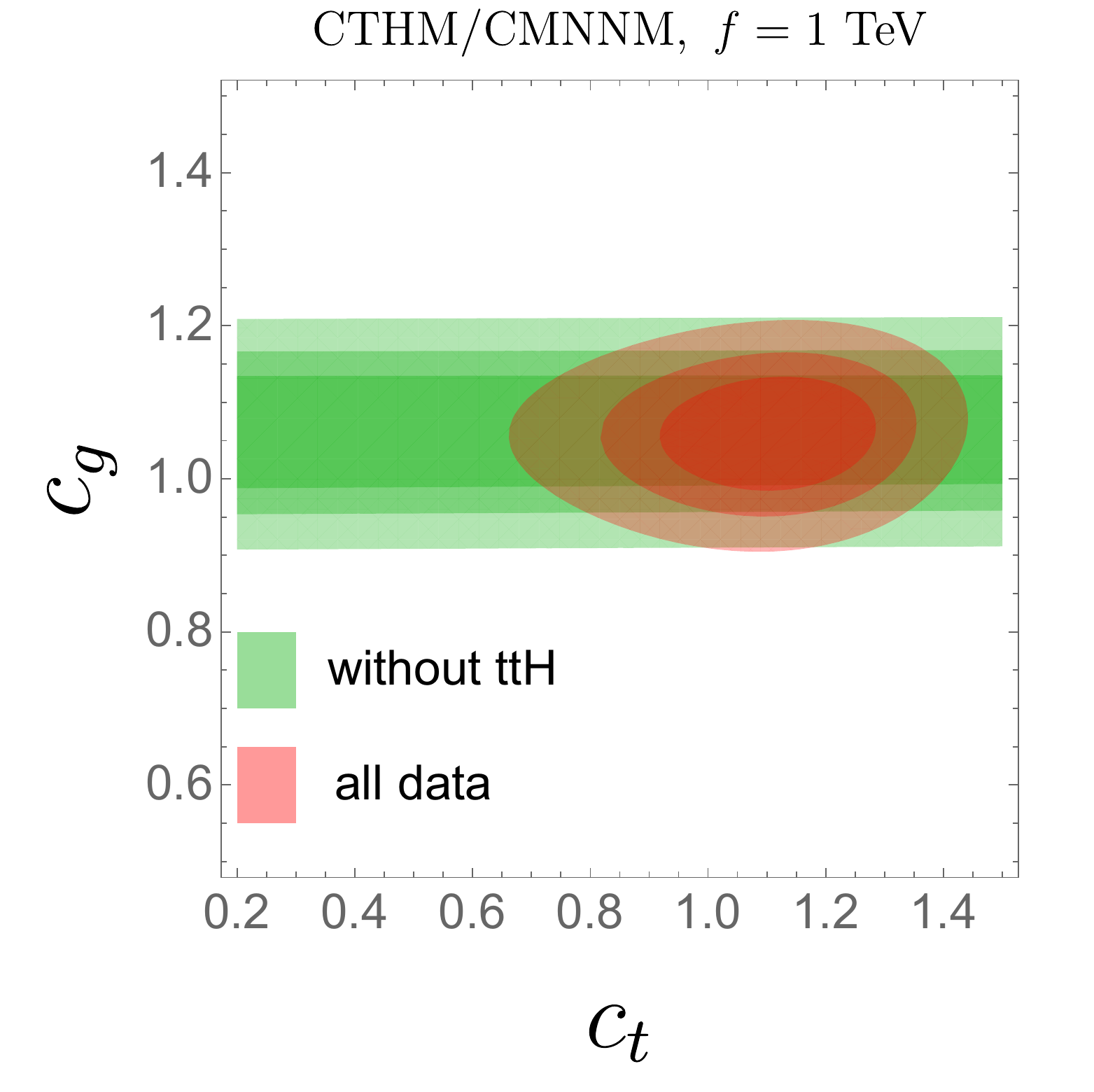}}
 \caption{The global fit on the values of $c_t$ and $c_g$ for MCHM (left) and CTHM/CMNNM (right). The green regions (from dark to light) are the $1\sigma$, $2\sigma$, and $3\sigma$ bounds without the $tth$ measurements corresponding to $\chi^2<(2.9,5.99,11.83)$. The red regions (from dark to light) are the $1\sigma$, $2\sigma$, and $3\sigma$ allowed regions corresponding to $\chi^2<(2.9,5.99,11.83)$ including all the Higgs signal measurements listed in the Table.~\ref{tab:atlas}\label{fig:ctcg_gf_2018}%
 }%
\end{figure}

The first set of experimental data we consider is the single Higgs measurement. We will perform a global fit analysis using the Higgs signal data listed in Tab.~\ref{tab:atlas}. From Sec.~\ref{sec:coupling}, we find that once we fix the global symmetry breaking scale $f$, the value of $c_g$ and $c_t$ will uniquely determine the signal strengths of all the combinations of the single Higgs production and decay channels listed in Tab.~\ref{tab:atlas}. 
For Higgs couplings to $\tau\tau$ and $bb$, we only take into account the effect that comes from Higgs nonlinearity, i.e. assuming $c_b=c_\tau=c_W=\sqrt{1-\xi}$ in CTHMs/CMNNM and $c_b=c_\tau=(1-2\xi)/\sqrt{1-\xi}$, $c_W=\sqrt{1-\xi}$ in MCHMs, and neglect the composite states for the $b$ and $\tau$ sector.
\begin{table}[ht]
\caption{Higgs Signal Measurements Used in The Global Fit}
\begin{center}
\begin{tabular}{cccccc}
    \hline
    \hline
    \multicolumn{6}{c}{ATLAS}\\
    \hline
          & $\gamma\gamma$ &$\tau\tau$ & $WW$ & $ZZ$&  $bb$\\
          \hline
     $ggH$& $0.81^{+0.19}_{-0.18}$\cite{Aaboud:2018xdt} &$1.02^{+0.63}_{-0.55}$\cite{Aaboud:2018pen} & $0.829^{+0.148}_{-0.142}$\cite{Aaboud:2018jqu} & $1.11^{+0.249}_{-0.225}$ \cite{Aaboud:2017vzb}     & N.A.\\
     $VBF$& $2.0^{+0.6}_{-0.5}$\cite{Aaboud:2018xdt} &$1.18^{+0.60}_{-0.54}$\cite{Aaboud:2018pen} & $1.626^{+0.977}_{-0.951}$\cite{Aaboud:2018jqu} & $3.987^{+1.728}_{-1.513}$ \cite{Aaboud:2017vzb}      & N.A.\\
     $VH$& $0.7^{+0.9}_{-0.8}$\cite{Aaboud:2018xdt} &N.A. & N.A. & N.A.     & $1.08^{+0.47}_{-0.43}(WH)$ $1.2^{+0.33}_{-0.31}$(ZH)\cite{Aaboud:2018zhk}\\
     $ttH$& $1.39^{+0.48}_{-0.42}$\cite{Aaboud:2018urx} &N.A. &N.A.& N.A.      &$0.79^{+0.61}_{-0.6}$\cite{Aaboud:2018urx}\\
    \hline
    \hline
\end{tabular}
\end{center}
\begin{center}
\label{tab:atlas}
\begin{tabular}{cccccc}
    \hline
    \hline
    \multicolumn{6}{c}{CMS\cite{Sirunyan:2018koj}}\\
    \hline
          & $\gamma\gamma$ &$\tau\tau$ & $WW$ & $ZZ$&  $bb$\\
          \hline
     $ggH$& $1.16^{+0.21}_{-0.18}$ &$1.05^{+0.53}_{-0.47}$ & $1.35^{+0.21}_{-0.19}$ & $1.22^{0.23}_{-0.21}$      & N.A.\\
     $VBF$& $0.67^{+0.59}_{-0.46}$ &$1.12^{+0.45}_{-0.43}$ & $0.28^{+0.64}_{-0.60}$ & $-0.09^{+1.02}_{-0.76}$       & N.A.\\
     $WH$& $3.76^{+1.48}_{-1.35}$ &N.A. & $3.91^{+2.26}_{-2.01}$ & $0.00^{+2.33}_{-0.00}$     & $1.73^{+0.7}_{-0.68}$\\
     $ZH$& $0.00^{+1.44}_{-0.00}$ &N.A. & $0.96^{+1.81}_{-1.46}$ & $0.00^{+4.26}_{-0.00}$     & $0.99^{+0.47}_{-0.45}$\\
     $ttH$& $2.18^{+0.88}_{-0.75}$ &$0.23^{+1.03}_{-0.88}$ & $1.60^{+0.65}_{-0.59}$ & $0.00^{+1.5}_{-0.00}$      &$0.91^{+0.45}_{-0.43}$\\
    \hline
    \hline
\end{tabular}
\end{center}
\end{table}
We therefore choose $c_g$ and $c_t$ as two independent parameters and perform a global fit for MCHM and CTHM/CMNNM independently. The results for $f=1$ TeV are shown in Fig.~\ref{fig:ctcg_gf_2018}, where the green bands represents the $1\sigma$, $2\sigma$, and $3\sigma$ bounds without taking into account the recent $tth$ measurements, while the red regions are those obtained with all the data listed in Tab.~\ref{tab:atlas}. The entry with N.A. in the table means the data is not currently available. We find that the global fit results are very similar within two scenarios, since the main difference comes from the Higgs couplings to bottom and $\tau$ leptons which is proportional to a small $\xi$.   

The details of the global fit are described below.
We use the public code \texttt{Lilith}~\cite{Bernon:2015hsa} to implement the global fit. We use the relative signal strength $\mu_{X,Y}$ defined below as observable: 
\begin{eqnarray}
\mu_{X,Y} = \frac{\sigma({X}\to H)BR(H\to Y)}{\sigma^{SM}({X}\to H)BR^{SM}(H\to Y)},
\end{eqnarray}
where $X$ represents the production mode, e.g. gluon fusion, vector boson fusion etc. and $Y$ represents the final state that the Higgs boson decays into. The test statistic $\chi^2$ is then constructed by:
\begin{eqnarray}
\chi^2=(\mu-\mu^{obs})^TC^{-1}(\mu-\mu^{obs}),
\end{eqnarray}
where $C^{-1}$ is the inverse of the covariance matrix $cov[\mu^{obs}_i,\mu^{obs}_j]$. In principle we need to know the whole $n\times n$ covariance matrix ($n$ is the number of observables we use in the global fit) to compute $\chi^2$, but this is obviously impossible and the relevant information is not provided by ATLAS and CMS collaborations. Therefore we just ignore the off-diagonal part in the covariance matrix and approximate the $\chi^2$ as:
\begin{eqnarray}
\chi^2=\sum_{X,Y}\frac{(\mu_{X,Y}-\mu^{obs}_{X,Y})^2}{\sigma^2_{X,Y}},
\end{eqnarray} 
where $\sigma_{X,Y}$ is the corresponding $1\sigma$ uncertainty for the given observable. For the detailed treatment of different plus and minus uncertainties one can consult the \texttt{Lilith} documentation~\cite{Bernon:2015hsa}.

Electroweak precision data (EWPD) is another set of experimental data that we use to constraint these models. A set of electroweak precision observable (EWPO) $S$, $T$, $W$, $Y$~\cite{Barbieri:2004qk} as an extension of the Peskin-Takeuchi parameters~\cite{Peskin:1991sw} can be defined to analyze the corrections coming from the heavy new physics under the assumption of the quark and lepton universality. Several detailed analysis of these observables in the minimal composite Higgs models and composite twin Higgs models can be found in Ref.~\cite{Grojean:2013qca,Barbieri:2007bh,Contino:2017moj}.
Due to the fact that the twin sector does not contribute to the EWPO at 1-loop level, the constraints for the MCHM and CTHM are similar. For simplicity, in our analysis we only take into account the constraint from the $T$ parameter with heavy composite fermions circulating in the loop, and approximate the contribution from the heavy resonance by the formula~\cite{Giudice:2007fh,Grojean:2013qca,Contino:2017moj}:
\begin{eqnarray}
T\sim \frac{3\xi}{16\pi^2}\frac{y_L^4f^2}{m_{min}^2},\ 
\label{eq:Tparameter}
\end{eqnarray} 
where the $m_{min}$ is the smallest mass parameter for the vector-like fermion resonance.

In addition to the above two sets of data, we also roughly take into account the constraint from the direct searches for top partners at the LHC~\cite{CMS:2019qjz,Aaboud:2018pii}. Depending on the dominant decay channel, the top partners mass has already been excluded up to around $1$ TeV to $1.3$ TeV. Therefore, in our parameter scan discussed below we set the minimum value of the mass parameters of those vector-like top partners to be $1$ TeV, which corresponds to larger value for the physical mass of the top partners.

\section{Numerical Analysis}
\label{sec:num}
\subsection{Parameter Scan}
To estimate the viable parameter space of each model under current experimental constraints we perform parameter scans with details explained as follows.
With the scale $f$ being fixed as $1$ TeV, we scan the parameter $y_L$ uniformly ranged between $-10$ to $10$. All the other dimensional parameters are scanned uniformly in a range from $1$ TeV to $10$ TeV. We afterward solve for the value of $y_R$ by requiring the mass of the top quark to be a value randomly chosen in a range from $150$ GeV to $170$ GeV. Finally we calculate the value of the effective couplings with the full expressions of the form factors in App.~\ref{sec:concreteff}. 
We then calculate the value of $T$ parameter using the approximate formula in Eq.~\ref{eq:Tparameter}, and only preserve points that satisfy the $T$ parameter constraint within $2\sigma$ level~\cite{Baak:2014ora}. 
We also put a rough requirement on the physical masses of top partners such that it is below the scale $4\pi f$, which is implemented by the following cuts:
\begin{eqnarray}
\sqrt{y_L^2f^2+m_{max}^2}<4\pi f{\ \rm and}\  \sqrt{y_R^2f^2+m_{max}^2}<4\pi f,
\end{eqnarray} 
where $m_{max}$ represents the largest mass parameter for the vector-like fermion resonance.

\subsection{Results}

Now we are ready to see what information we can extract with parameter scans.

\begin{figure}[htp]
{\includegraphics[width=0.45\textwidth]{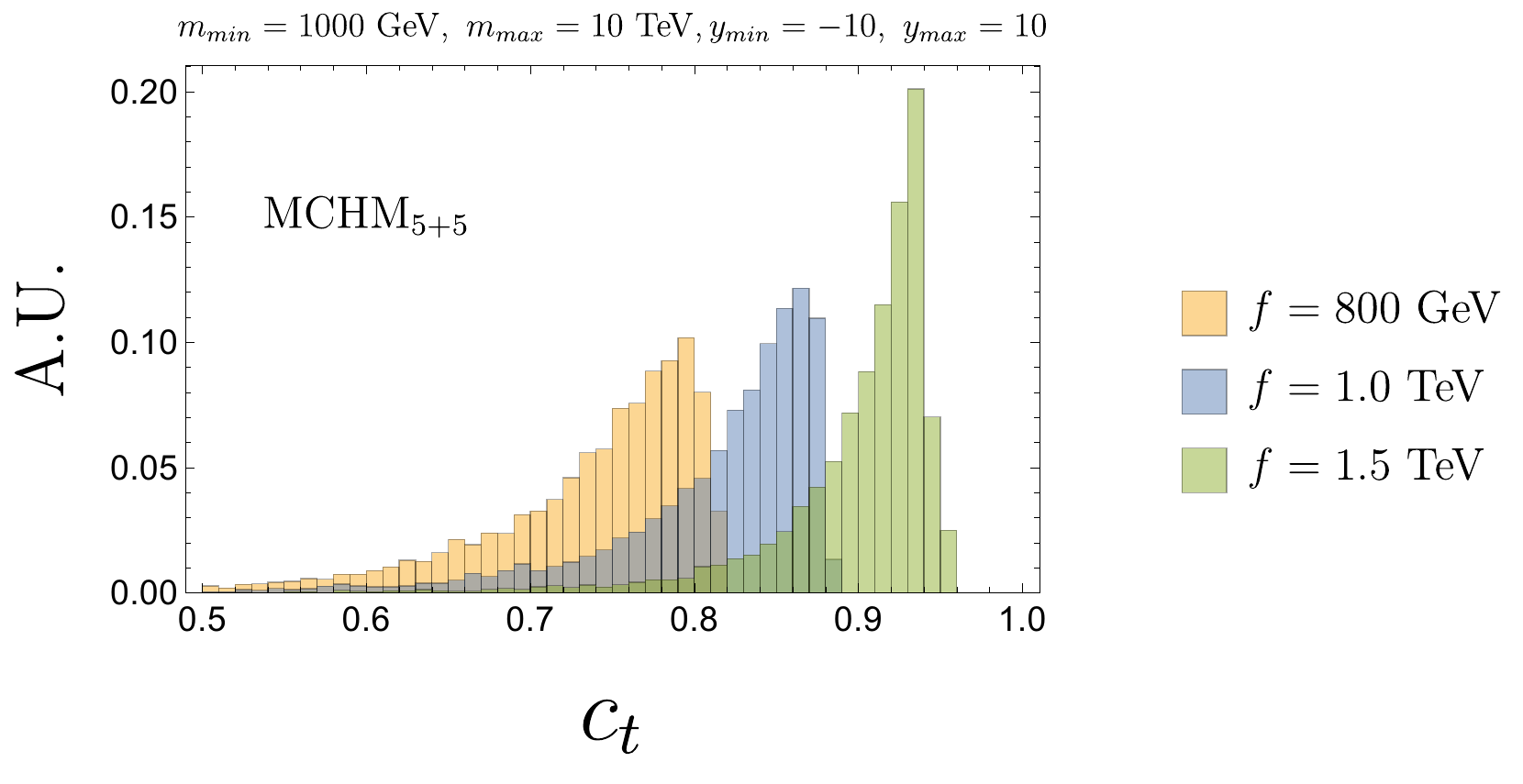}}%
 \quad
{\includegraphics[width=0.45\textwidth]{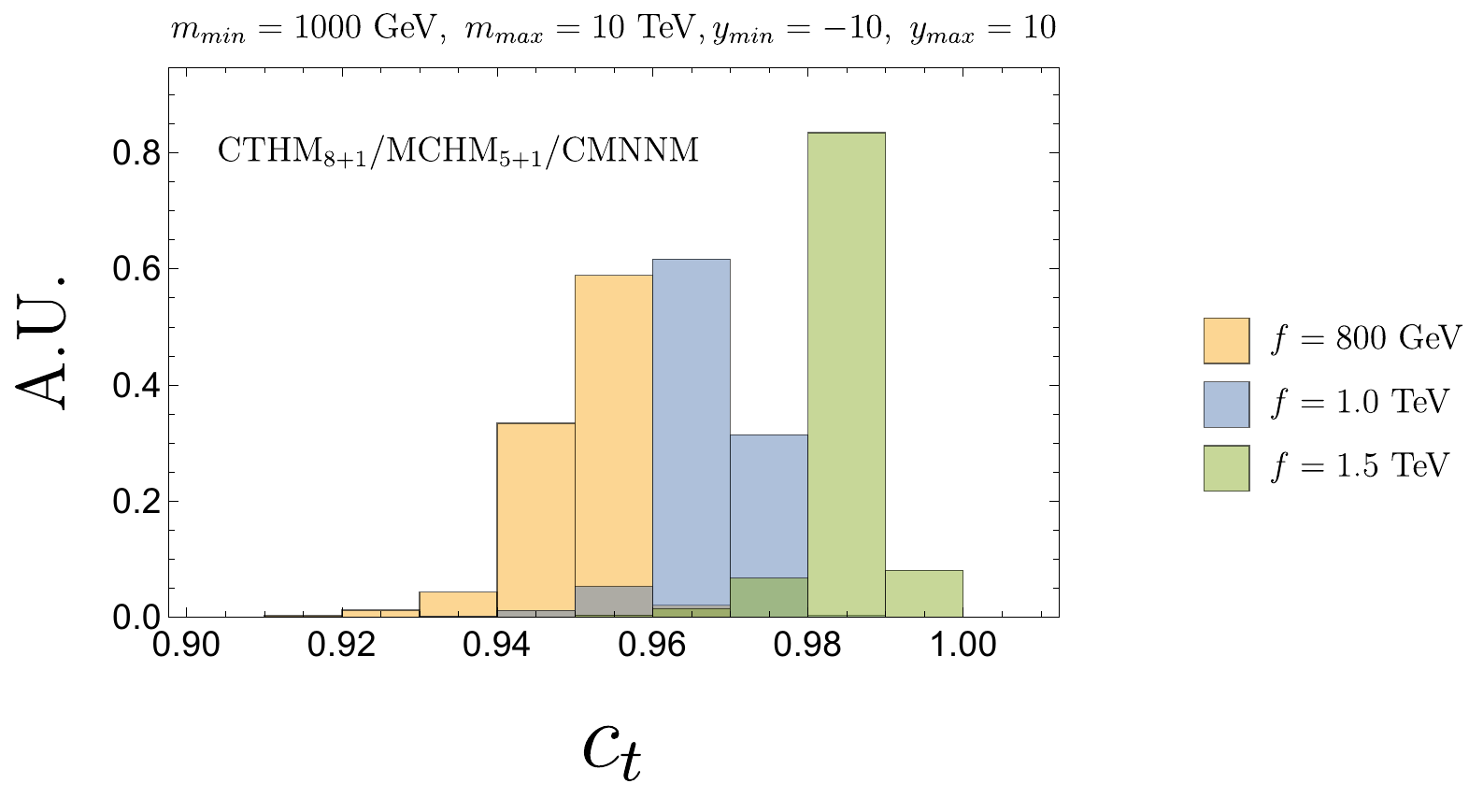}}\\
{\includegraphics[width=0.45\textwidth]{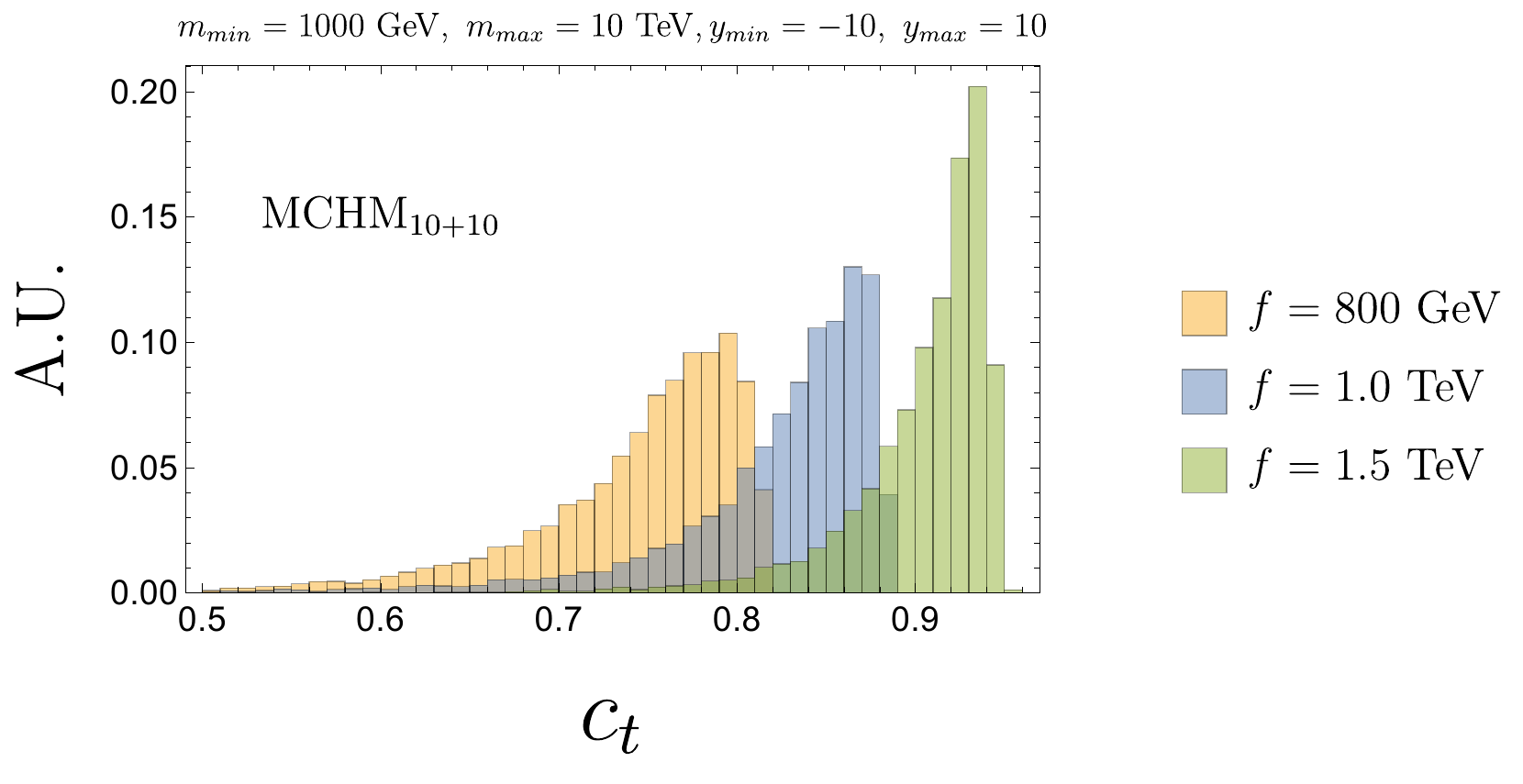}}%
 \quad
{\includegraphics[width=0.45\textwidth]{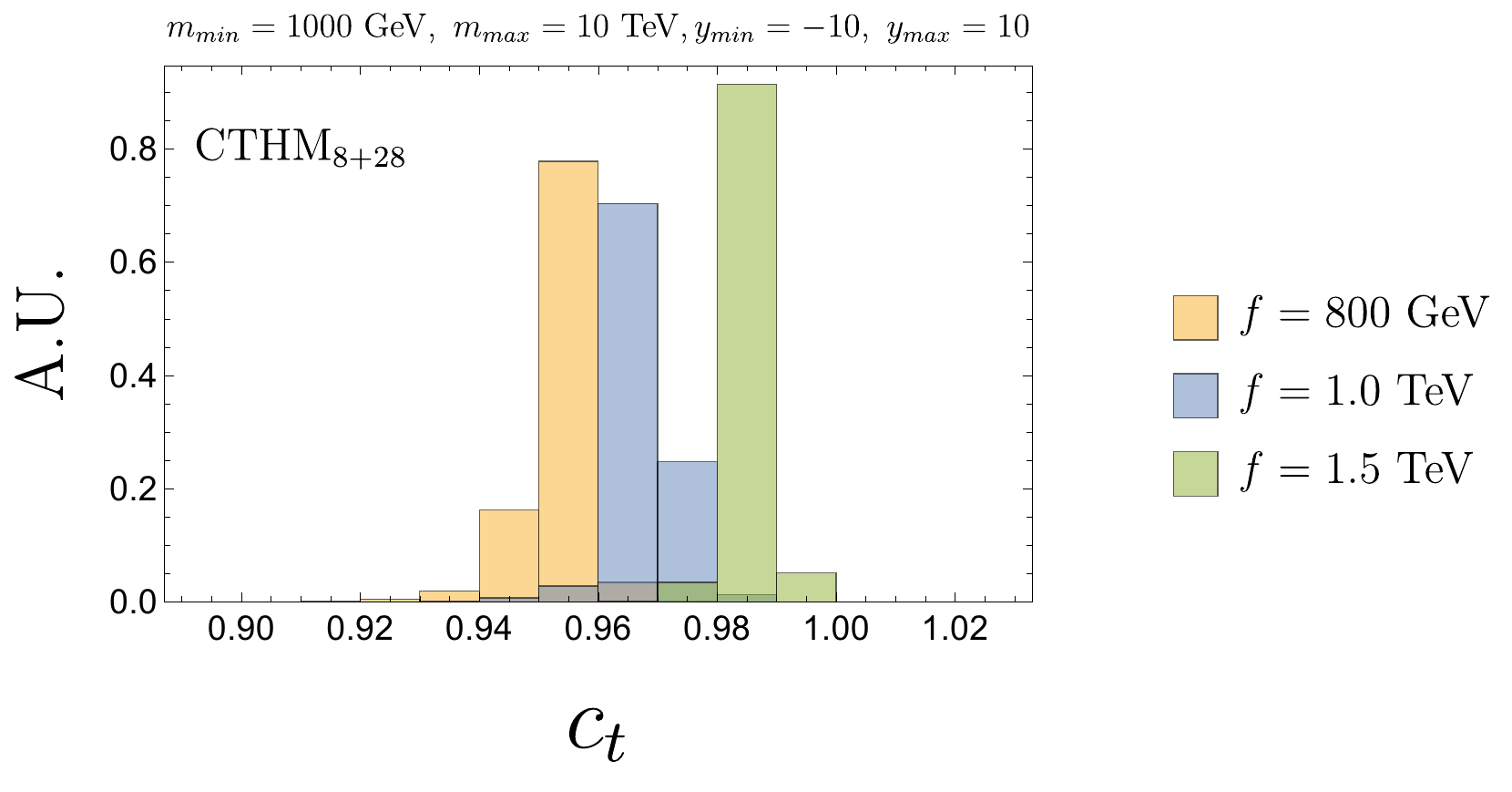}}
 \caption{The distribution of $c_t$ in the models with low dimensional representations. Different colors denote different values of $\xi$, the width of each bin is chosen as $0.01$, and A.U. denotes arbitrary unit.\label{fig:ct_low}%
 }%
\end{figure}

\begin{figure}[htp]
{\includegraphics[width=0.45\textwidth]{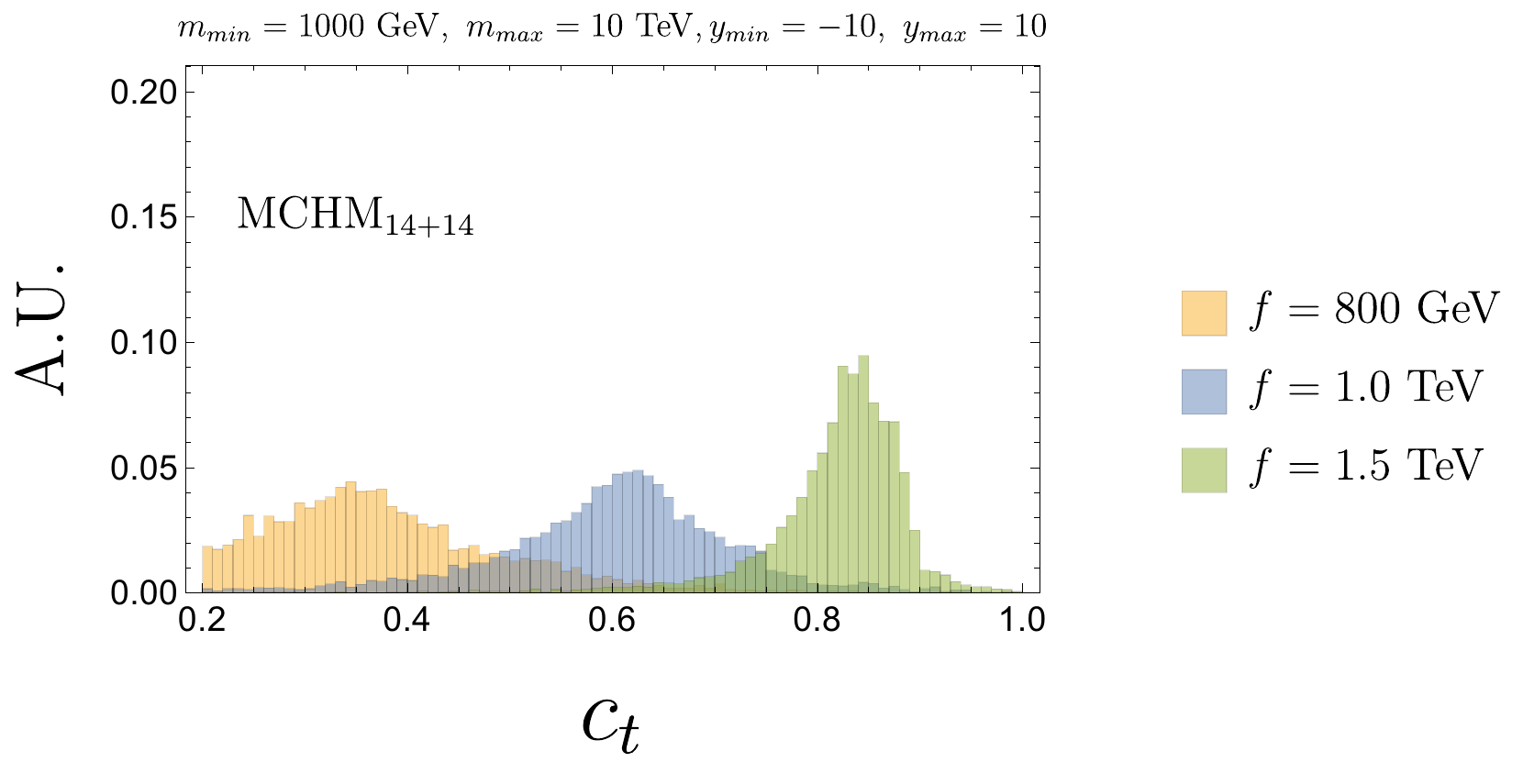}}%
 \quad
{\includegraphics[width=0.45\textwidth]{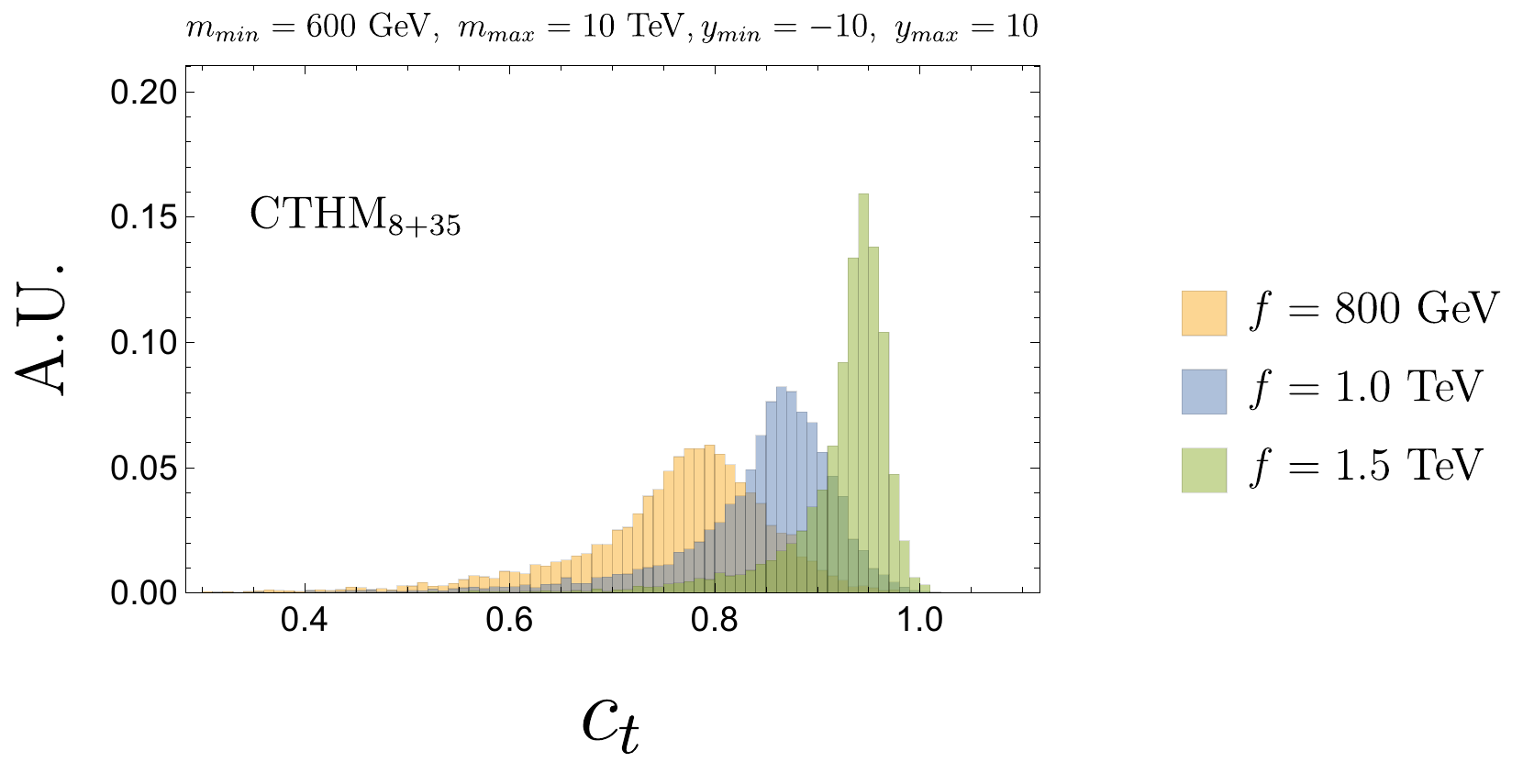}}
 \caption{The distribution of $c_t$ in the models with low dimensional representations. Different colors denote different values of $\xi$, the width of each bin is chosen as $0.01$, and A.U. denotes arbitrary unit.
\label{fig:ct_high}%
 }%
\end{figure}

Firstly, we present the results of distribution of $c_t$ in each model and see the effect of the value of $f$ on these distributions. Fig.~\ref{fig:ct_low} and Fig.~\ref{fig:ct_high} are the plots of the distributions for the models with low and high dimensional fermion representations respectively. More specifically, the low dimensional representations refer to MCHM$_{5+1,5+5,10+10}$ and CTHM$_{8+1,8+28}$. We put the distribution MCHM$_{5+1}$ and CTHM$_{8+1}$ in the same plot, since the expressions of form factors are the same in these two models. We find following features from these plots:

\begin{itemize}
\item The peak of $c_t$ shifts downwards as the global symmetry breaking scale $f$ decreases. This can be understood by observing the expression for $c_t$ in Eq.~\ref{eq:ctmchm} and \ref{eq:ctcthm}. The value of $\xi$ determine the overall magnitude of the deviation from one.
\item In the low dimensional representations, the spans of the $c_t$ in the CTHMs are much smaller than those in the MCHMs. The reason is that the form factor $\Pi_{1t_Lt_R}$ depends on two mass parameters in the MCHM, while it depends on only one mass parameter in the CTHM (To be specific, $m_1$ in CTHM$_{8+1}$ and $m_7$ in CTHM$_{8+28}$ as shown in App.~\ref{sec:concreteff}).
Therefore, less freedom in the parameter space is left for the CTHM-type of models to tune the parameters to reproduce the top quark mass. 
\end{itemize}

\begin{figure}[htp]
{\includegraphics[width=0.35\textwidth]{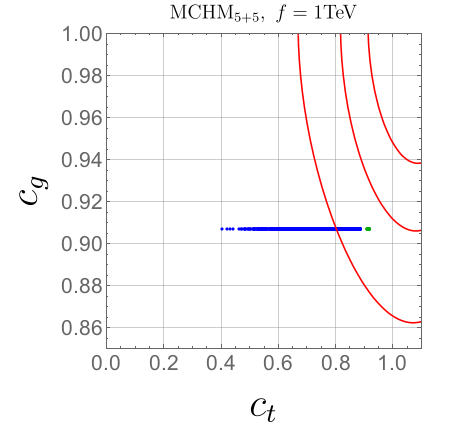}}%
 \quad
{\includegraphics[width=0.35\textwidth]{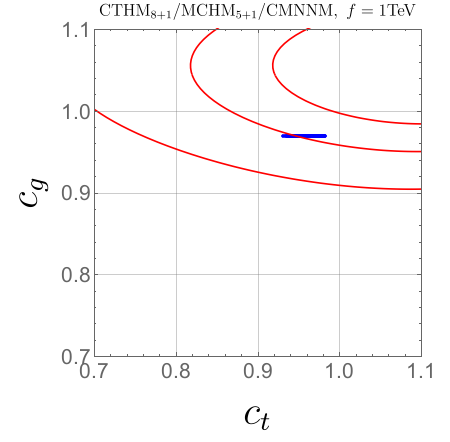}}\\
{\includegraphics[width=0.35\textwidth]{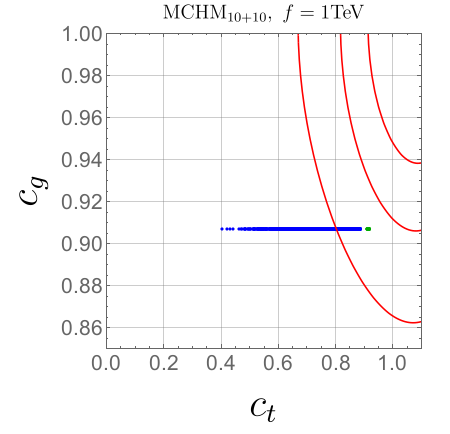}}%
 \quad
{\includegraphics[width=0.35\textwidth]{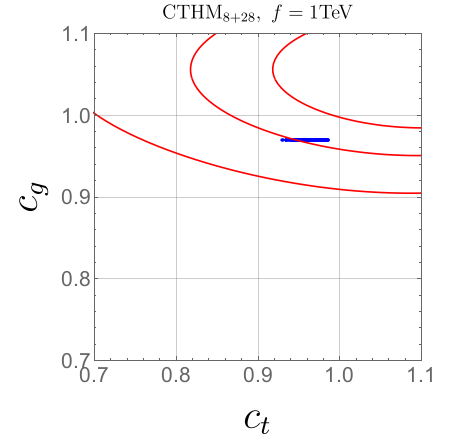}}
 \caption{\small The distribution of $c_t$ vs $c_g$ in models with low dimensional representations. The scale $f$ is set to $1$ TeV. The colored lines are $1\sigma$, $2\sigma$ and $3\sigma$ bounds coming from the Higgs signal global fit using Run2 data. In MCHM, the green points predict $c_t>c_g$ thus the model has the problem of triggerring EWSB~\cite{Liu:2017dsz}, while the blue points satisfy $c_g>c_t$. In CTHMs, EWSB is automatically triggered as shown in Sec.~\ref{subsec:pot}.\label{fig:ctcg_gf1}%
 }%
\end{figure} 

\begin{figure}[htp]
{\includegraphics[width=0.35\textwidth]{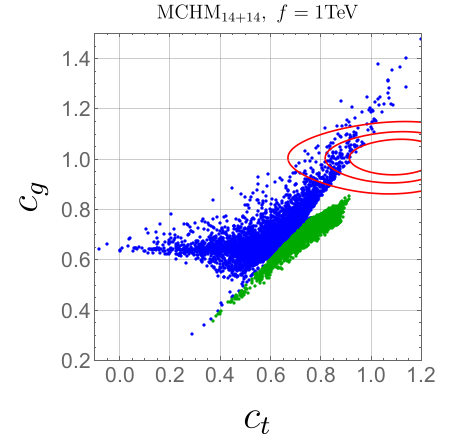}}%
 \quad
{\includegraphics[width=0.35\textwidth]{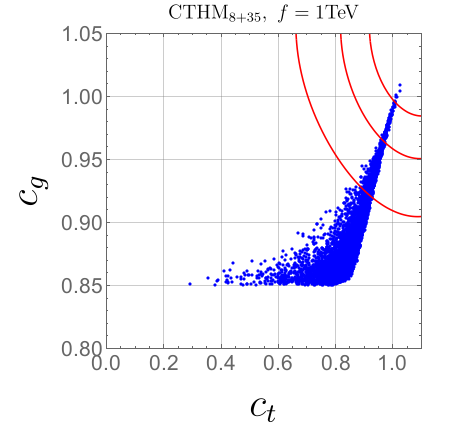}}
 \caption{\small The distribution of $c_t$ vs $c_g$ in models with high dimensional representations. The scale $f$ is set to $1$ TeV. The colored lines are $1\sigma$, $2\sigma$ and $3\sigma$ bounds coming from the Higgs signal global fit using Run2 data. In the MCHM, the green points predict $c_t>c_g$ thus the model has the problem of triggering EWSB~\cite{Liu:2017dsz}, while the blue points satisfy $c_g>c_t$. In CTHMs, EWSB is automatically triggered as shown in Sec.~\ref{subsec:pot}.\label{fig:ctcg_gf2}
 }%
\end{figure} 

Secondly, we analyze the viable parameter region of each model under current experimental constraints taking into account the results of the global fit on $c_g$ vs $c_t$ plane. In the following analysis we focus on the benchmark value $f=1$ TeV. In Fig.~\ref{fig:ctcg_gf1} and~\ref{fig:ctcg_gf2}, we overlap the $1\sigma$, $2\sigma$, and $3\sigma$ contours from our global fit to the parameter scan in $c_t$ vs $c_g$ plane. The green dots in the MCHMs predict $c_t>c_g$, thus the model may suffer from the problem of the non-existence of EWSB~\cite{Liu:2017dsz}. However, EWSB is automatically triggered in CTHM-type of models as discussed in Sec.~\ref{subsec:pot}, so we did not separate the points with different colors. From these plots we can find the following facts:
\begin{itemize}
\item 
The new measurements of $tth$ production impose a strong constraint on the value of $c_t$ such that all the models are only moderately compatible with the global fit result if $f=1$ TeV. At the worst, MCHM with $5+5$ and $10+10$ representations are disfavored at the 2$\sigma$ confidence level (CL) for $f=1$ TeV. CTHM with $8+1$ and $8+28$ representations can have most points within the 2$\sigma$ region but outside the 1$\sigma$ region for $f=1$ TeV.
\item 
The high dimensional representations can roughly be more consistent with the global fit constraints than the low dimensional representations. Especially in the CTHM$_{8+35}$, the points within the 1$\sigma$ region is still possible for $f=1$ TeV. Moreover, if the future experimental result confirms that $c_t$ is preferred to be larger than $1$, then the parameter space in models with high dimensional representations is more available.
\item 
In the low dimensional representations, both values of $c_g$ and $c_W$ are fixed by the value of $\xi$, i.e. the global symmetry breaking scale $f$. In the future, if $\xi$ is obtained by the measurements of $c_W$ for example from $e^+e^-$ collider with Higgsstrahlung process, then one can check whether the measured value of $c_g$ agrees with the correlation of $c_g$ and $c_W$. The significant deviation from the correlation will disfavor low representations, or it can shed light on the extra heavy particles that explicitly break the shift symmetry of the PNGB Higgs~\cite{Cao:2018cms}.
\end{itemize}

\begin{figure}[htp]
{\includegraphics[width=0.35\textwidth]{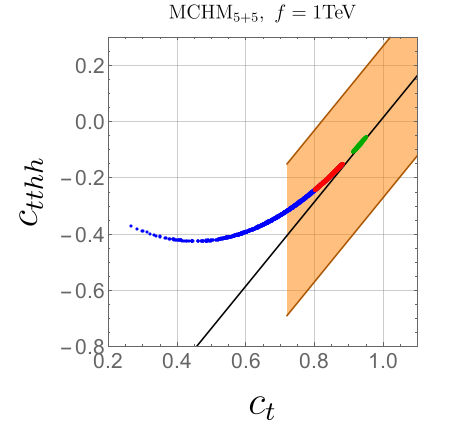}}%
 \quad
{\includegraphics[width=0.35\textwidth]{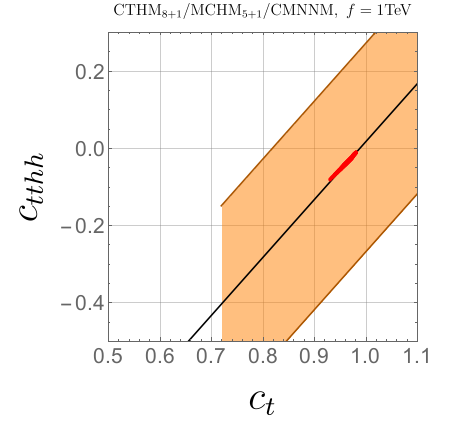}}\\
{\includegraphics[width=0.35\textwidth]{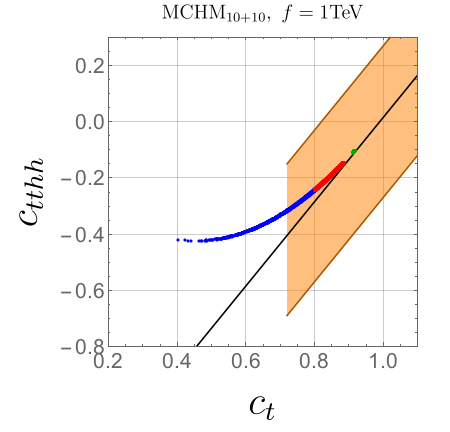}}%
 \quad
{\includegraphics[width=0.35\textwidth]{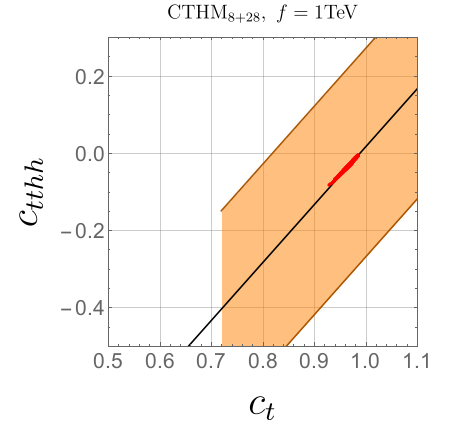}}
 \caption{The distribution of $c_{tthh}$ vs $c_t$ in models with low dimensional representations. The scale $f$ is set to 1 TeV. In the MCHM, the green points predict $c_t>c_g$ thus the model has the problem of triggering EWSB~\cite{Liu:2017dsz}, while the blue points satisfy $c_g>c_t$. EWSB is automatically triggered in CTHMs. The black line represents the relationship between $c_t$ and $c_{tthh}$ in Eq.~\ref{eq:ctthhctch}. The orange block represents the relation of Eq.~\ref{eq:ctthhctnscalar} with $c_W$ and $c_t$ within $2\sigma$ uncertainties from current Higgs signals~\cite{Aaboud:2018urx}. The red dots are within the 3$\sigma$ region of global fit as shown in Fig.~\ref{fig:ctcg_gf2}. }
\label{fig:ctctthh_gf1}
\end{figure} 
 
\begin{figure}[htp]
{\includegraphics[width=0.35\textwidth]{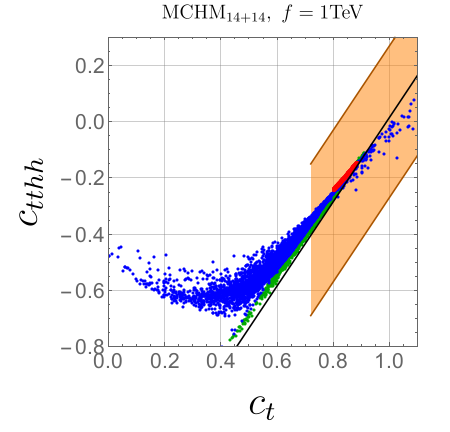}}%
 \quad
{\includegraphics[width=0.35\textwidth]{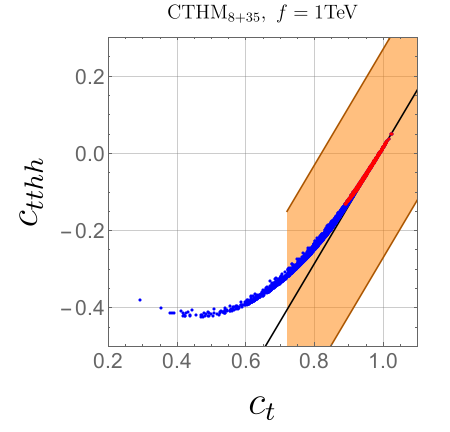}}
 \caption{The distribution of $c_{tthh}$ vs $c_t$ in models with high dimensional representations. The scale $f$ is set to 1 TeV. In the MCHM, the green points predict $c_t>c_g$ thus the model has the problem of triggering EWSB~\cite{Liu:2017dsz}, while the blue points satisfy $c_g>c_t$. EWSB is automatically triggered in CTHMs. The black line represents the relationship between $c_t$ and $c_{tthh}$ in Eq.~\ref{eq:ctthhctch}. The orange block represents the relation of Eq.~\ref{eq:ctthhctnscalar} with $c_W$ and $c_t$ within $2\sigma$ uncertainties from current Higgs signals~\cite{Aaboud:2018urx}. The red dots are within the 3$\sigma$ region of global fit as shown in Fig.~\ref{fig:ctcg_gf2}.}
\label{fig:ctctthh_gf2}
\end{figure} 
 
Thirdly, we investigate the correlation between $c_{tthh}$ and $c_t$ in MCHMs and CTHMs, and their interplay with the global fit.
In Fig.~\ref{fig:ctctthh_gf1} and ~\ref{fig:ctctthh_gf2}, we plot the points predicted by models on the $c_{tthh}$ vs $c_t$ plane. 
We use the black line in each plot to denote the relation between $c_{tthh}$ and $c_{t}$ when expanding with respect to $\xi$ to the linear order, i.e. Eq.~\ref{eq:ctthhctmchm} and \ref{eq:ctthhctcthm}. We reorganize it as  the following:
\begin{eqnarray}
c_{tthh}=\frac{3}{2}c_t-\frac{3}{2}+\frac{\xi}{4}.
\label{eq:ctthhctch}
\end{eqnarray} 
In the meantime, we also include the orange region based on the following formula obtained from the framework of dimension-six SMEFT~\cite{Corbett:2017ieo}:
\begin{equation}
c_{tthh}=\frac{3}{2}c_t-\frac{1}{2}c_W-1,
\label{eq:ctthhctnscalar}
\end{equation}
with both $c_W$ and $c_t$ within the $2\sigma$ region from the Higgs signal global fit in the $\kappa$ framework~\cite{Sirunyan:2018koj}. We emphasis here that Eq.~\ref{eq:ctthhctnscalar} is valid whether Higgs is fundamental or composite. 
The red dots that we highlighted in these plots are the points that satisfy the 3$\sigma$ global fit constraint taking into account the correlation between the Higgs effective couplings, i.e. the points inside the $3\sigma$ region (marked in red) in Fig.~\ref{fig:ctcg_gf_2018}.

Several comments are in order after combing the information from Eq.~\ref{eq:ctthhctch}, Eq.~\ref{eq:ctthhctnscalar} and Fig.~\ref{fig:ctctthh_gf1}, Fig.~\ref{fig:ctctthh_gf2}:
\begin{itemize}
\item 
If one plugs in the expression of $c_W=\sqrt{1-\xi}$ in composite Higgs models into Eq.~\ref{eq:ctthhctnscalar} and keep the linear term with the expansion of $\xi$, one can recover the relation of Eq.~\ref{eq:ctthhctch}. This indicate that if the linear approximation of $\xi$ is valid in the composite Higgs models, then one cannot use the relation in Eq.~\ref{eq:ctthhctch} to test the effect of Higgs nonlinearity.  
\item
The red dots, which are the parameters points within the $3\sigma$ global fit bound, are aligned with the linear approximation (black line) in CTHMs, thus Higgs nonlinearity cannot be tested through the relation in Eq.~\ref{eq:ctthhctnscalar} in this case. However, the Higgs nonlinearity effect can be shown in various MCHMs, i.e. the red dots in MCHMs are possible to have some deviation from the black line. 
\end{itemize}

\section{Conclusion}
\label{sec:con}
In this work, we focus on the top sector in several composite Higgs models (including hidden sectors) that can realize the naturalness conditions. 
We find that the quadratic divergence can be cancelled out by one of the following symmetries: collective symmetry, left-right $Z_2$ symmetry and the mirror $Z_2$ symmetry. Instead of working in any specific model, one can integrate out those composite top partners introduced for the naturalness requirement and utilize the general form factors to describe strong dynamics at TeV scale. We then systematically obtain the Higgs couplings with the top sector in the framework of minimal composite Higgs models and composite twin Higgs models, composite minimal neutral naturalness model, where the left-handed and right-handed top quark are embedded in different representations of the global symmetry. Both the Higgs nonlinearity as well as the compositeness from the top partners could induce the deviation of Higgs couplings from the SM values. 

Theoretically, pattern of the Higgs effective couplings is reflected by the Higgs dependence in the form factors. The Higgs dependence of the form factor $\Pi_{LR}$, the two point correlation function between the left-handed and right-handed top quarks, can be completely determined by symmetries, without the need of tedious calculation. We find in composite twin Higgs models $\Pi_{LR}$ satisfy a universal expansion as in Eq.~\ref{eq:twin} regardless of the specific fermion representations. 
This fact is dictated by the Higgs dependence constrained by the $Z_2$ symmetry of the twin Higgs setup. On the other hand, $\Pi_{LR}$ can satisfy different expansions of PNGB-Higgs dependence in minimal composite Higgs models depending on the choices of top-quark embeddings. 
These are new features presented in this paper.

Numerically, we perform  global fits on Higgs couplings and parameter scan in various models. We find the following conclusion in our study:
\begin{itemize}
\item We update the existing global fit of single Higgs measurements by including the latest $tth$ data, which starts to put constraint on $c_t$, and thus exclude further parameter space.  Current global fit of single Higgs measurements favor high dimensional representations in both minimal composite Higgs and composite twin Higgs models, which predict $c_t$ could be larger than one. If future measurements confirm an enhanced $ttH$ coupling, then low dimensional representations will be disfavored in the both minimal composite Higgs and composite twin Higgs models for $f=1$ TeV.
\item The impact of Higgs nonlinearity effect on effective Higgs couplings is enhanced if composite particles in the spectrum have significant mass splittings, caused by the mass difference of full composite multiplets as well as the mixing between components inside individual composite multiplet and the elementary fermions. As a result, certain combination of the form factors can cause the terms proportional to $\xi^2$, or higher powers, being at the same of order of the ones proportional to $\xi$. 
\item There are two interesting correlations: $c_W$ verses $c_g$, and $c_{tthh}$ verses $c_{t}$. The first correlation can be very strong in low dimensional representations. Thus if such correlation is not observed, then the top quark is favored to belong to high dimensional fermion representation.
If the second correlation is violated then MCHM is favored, as one can see from the plots that the red dots are mostly aligned with the black line in CTHMs in Fig~\ref{fig:ctctthh_gf1} and \ref{fig:ctctthh_gf2}.
\end{itemize}
Overall, precise measurements of various Higgs couplings at future colliders will help us to discriminate the nature of the Higgs boson, the fermion embeddings, and eventually the origin of the electroweak symmetry breaking. 

\begin{acknowledgements}
We thank Qing-Hong Cao, Jing Shu and Bin Yan for useful discussions.
H.L.L. and J.H.Y. are supported by the National Science Foundation of China under Grants No. 11875003.
L.X.X. and S.H.Z. are supported in part by the National Science Foundation of China under Grants No. 11635001, 11875072. J.H.Y. is also supported by the Chinese Academy of Sciences (CAS) Hundred-Talent Program.
\end{acknowledgements}

\appendix
\label{sec:app}

\section{Form Factors in the Bosonic Sector}
\label{app:bosonic}
Within the Landau gauge $\partial_\mu A^{a\mu\nu}=0$, the general Lagrangian describing the bosonic sector of composite Higgs can be written as~\cite{Contino:2010rs} (see e.g. Ref.~\cite{Marzocca:2012zn} for study in details.)
\bea
\mathcal{L}_{\textnormal{boson}}=\frac{1}{2}(P_T)^{\mu\nu}\left[\Pi_0(q^2)\textnormal{Tr}(A_\mu A_\nu)+\Pi_1(q^2)\Sigma^\dagger A_\mu A_\nu \Sigma\right]
\eea
up to the quadratic level of gauge bosons in the momentum space.
Here $A_\mu\equiv A^a_\mu T^a$ where $A^a_\mu$ denote the gauge bosons associated with the corresponding generators of the broken global symmetry group $G$. $\Sigma$ are goldstone bosons of the coset $G/H$. $P_T^{\mu\nu}$ is the projection operator
\bea
P_T^{\mu\nu}=\eta^{\mu\nu}-\frac{q^\mu q^\nu}{q^2}.
\eea
The extra $U(1)_X$ gauge boson, which is usually necessary to reproduce correct fermion hyper-charges, has been neglected in the above Lagrangian.

For cosets in which we are interested in this paper, $\Sigma$ are explicitly
\bea
\begin{aligned}
\Sigma&=(0,0,0,s_h,c_h)^T\ \ \ \ \ \ \ \ \ \ \ \ \ \ SO(5)/SO(4)\\
\Sigma&=(0,0,0,s_h,0,0,0,c_h)^T\ \ \ \ \ \ SO(8)/SO(7)\\
\end{aligned}
\eea
in the unitary gauge. With the form factors at the limit $Q^2\to 0$ as $\Pi_0(0)=0$ and $\Pi_1(0)=f^2$, 
we read off the Higgs-dependent $W^{\pm}$ boson mass directly
\bea
m_W^2(h)=\frac{g^2}{4}v^2=\frac{g^2f^2}{4}\sin^2\left(\frac{h}{f}\right), 
\eea
from which the Higgs coupling to electroweak gauge bosons $c_W$ is derived.

\section{More on Higgs Effective Couplings}
The relevant dimension-six operators are
\begin{align}
\mathcal{L}_{D=6}&=\frac{\mathbb{C}_H}{2f^2}\partial^\mu (H^\dagger H)\partial_\mu (H^\dagger H) + \frac{\mathbb{C}_T}{2f^2}\left (H^\dagger {\overleftrightarrow{D}_\mu} H\right)^2 -\frac{\mathbb{C}_6 \lambda}{f^2}(H^\dagger H)^3\nn\\
&+\left(\frac{\mathbb{C}_y y_f}{f^2}H^\dagger H \bar{f}_LH f_R+\text{h.c.}\right)+\frac{\mathbb{C}_g g_s^2}{16 \pi^2 f^2}\frac{y_t^2}{g_\rho^2}H^\dagger H G^a_{\mu\nu} G^{a\mu\nu}+\frac{\mathbb{C}_\gamma g^{\prime 2}}{16 \pi^2 f^2}\frac{g^2}{g_\rho^2}H^\dagger H B_{\mu\nu} B^{\mu\nu}\,
\label{eq:silh}
\end{align}
where $\mathbb{C}_{H,T,6,y,g,\gamma}$ are the unknown Wilson coefficients.
The operator with coefficient $\mathbb{C}_T$ violates custodial symmetry at tree level and is tightly constrained by precision electroweak data, so we can ignore it.
The new physics scale and the typical coupling strength of the UV theory are denoted as $f$ and $g_\rho$ respectively. 

One can also match the Wilson coefficients in Eq.~\ref{eq:silh} with the general form factors of composite Higgs models.
For minimal composite Higgs models, we have
\begin{align}
\mathbb{C}_H&=\frac{2}{\xi}(1-c_W)=1+\mathcal{O}(\xi)\ ,\nn\\
\mathbb{C}_y&=\frac{1}{\xi}(1-c_t)-\frac{c_H}{2}=1+\left(\frac{\Pi_{1t_L}(0)}{\Pi_{0t_L}(0)}+\frac{\Pi_{1t_R}(0)}{\Pi_{0t_R}(0)}\right)-2 \frac{\Pi_{2t_Lt_R}}{\Pi_{1t_Lt_R}}+\mathcal{O}(\xi)\ ,\nn\\
\mathbb{C}_6&=0+\mathcal{O}(\xi)\ ,\nn\\
\mathbb{C}_g&=\frac{g_\rho^2}{3y_t^2}\frac{1}{\xi}(c_g-c_t)=\frac{g_\rho^2}{3y_t^2}\left(\frac{\Pi_{1t_L}(0)}{\Pi_{0t_L}(0)}+\frac{\Pi_{1t_R}(0)}{\Pi_{0t_R}(0)}\right)+\mathcal{O}(\xi)\ ;
\end{align}
for composite twin Higgs models, we have
\begin{align}
\mathbb{C}_H&=\frac{2}{\xi}(1-c_W)=1+\mathcal{O}(\xi)\ ,\nn\\
\mathbb{C}_y&=\frac{1}{\xi}(1-c_t)-\frac{c_H}{2}=\left(\frac{\Pi_{1t_L}(0)}{\Pi_{0t_L}(0)}+\frac{\Pi_{1t_R}(0)}{\Pi_{0t_R}(0)}\right)-2 \frac{\Pi_{2t_Lt_R}}{\Pi_{1t_Lt_R}}+\mathcal{O}(\xi)\ ,\nn\\
\mathbb{C}_6&=0+\mathcal{O}(\xi)\ ,\nn\\
\mathbb{C}_g&=\frac{g_\rho^2}{3y_t^2}\frac{1}{\xi}(c_g-c_t)=\frac{g_\rho^2}{3y_t^2}\left(\frac{\Pi_{1t_L}(0)}{\Pi_{0t_L}(0)}+\frac{\Pi_{1t_R}(0)}{\Pi_{0t_R}(0)}\right)+\mathcal{O}(\xi)\ .
\end{align}
\section{Form Factors in Specific Composite Models}
\label{sec:concreteff}
In this part, we present the form factors in specific minimal composite Higgs models and composite twin Higgs models. To avoid confusion, we explicitly present the Higgs dependence in the chirality-flipped form factor for $\textrm{MCHM}_{5+1}$, $\textrm{MCHM}_{14+1}$ and $\textrm{CMNNM}$.

\begin{itemize}
\item $\textnormal{MCHM}_{5+1}$:\\
\bea
\begin{aligned}
\mathcal{L}_{5+1}&=y_Lf (\bar{q}^5_L)^i \left[U_{iJ} \Psi_4^J+U_{i5}\Psi_1\right]+y_Rf\bar{t}_R\Psi_{1L}+\textnormal{h.c.}-m_4\bar{\Psi}_4\Psi_4-m_1\bar{\Psi}_1\Psi_1\\
\end{aligned}
\eea
\bea
\begin{aligned}
\Pi_{0t_L} &=1-\frac{y^2_Lf^2}{p^2-m_4^2}\\
\Pi_{1t_L} &=\frac{y_L^2f^2}{2}\left(\frac{1}{p^2-m^2_4}-\frac{1}{p^2-m^2_1}\right)\\
\Pi_{0t_R} &=1-\frac{y^2_Rf^2}{p^2-m_1^2}\\
\Pi_{t_Lt_R} &=-\frac{m_1}{\sqrt{2}}\cdot\frac{y_Ly_Rf^2}{p^2-m^2_1}s_h\\
\end{aligned}
\eea

\item $\textnormal{MCHM}_{5+5}$:\\
\bea
\begin{aligned}
\mathcal{L}_{5+5}&=y_Lf (\bar{q}^5_L)^i \left[U_{iJ} \Psi_4^J+U_{i5}\Psi_1\right]+y_Rf (\bar{t}^5_R)^i \left[U_{iJ} \Psi_4^J+U_{i5}\Psi_1\right]+\textnormal{h.c.}\\
&\ \ \ \ -m_4\bar{\Psi}_4\Psi_4-m_1\bar{\Psi}_1\Psi_1
\end{aligned}
\eea
\bea
\begin{aligned}
\Pi_{0t_L}&=1-\frac{y_L^2f^2}{p^2-m_4^2}\\
\Pi_{1t_L}&=\frac{y_L^2f^2}{2}\left(\frac{1}{p^2-m_4^2}-\frac{1}{p^2-m_1^2}\right)\\
\Pi_{0t_R}&=1-\frac{y_R^2f^2}{p^2-m_1^2}\\
\Pi_{1t_R}&=y_R^2f^2\left(\frac{1}{p^2-m_1^2}-\frac{1}{p^2-m_4^2}\right)\\
\Pi_{1t_Lt_R}&=\frac{1}{\sqrt{2}}y_Ly_Rf^2\left(\frac{m_4}{p^2-m_4^2}-\frac{m_1}{p^2-m_1^2}\right)
\end{aligned}
\eea

\item $\textnormal{MCHM}_{10+10}$:\\
\bea
\begin{aligned}
\mathcal{L}_{10+10}&=y_L f (\bar{q}^{10}_L)^{ij}\left[U_{jJ}U_{iL}\Psi^{JL}_{6}+\sqrt{2}\ U_{i5}U_{jJ}\Psi^{J}_{4}\right]\\
&\ \ +y_R f (\bar{t}^{10}_R)^{ij}\left[U_{jJ}U_{iL}\Psi^{JL}_{6}+\sqrt{2}\ U_{i5}U_{jJ}\Psi^{J}_{4}\right]+\textnormal{h.c.}\\
&\ \ -m_{6}\bar{\Psi}_{6}\Psi_{6}-m_4\bar{\Psi}_{4}\Psi_{4}
\end{aligned}
\eea 
\bea
\begin{aligned}
\Pi_{0t_L}&=1-\frac{y_L^2f^2}{p^2-m_4^2}\\
\Pi_{1t_L}&=\frac{y_L^2f^2}{2}\left(\frac{1}{p^2-m_4^2}-\frac{1}{p^2-m_6^2}\right)\\
\Pi_{0t_R}&=1-\frac{y_R^2f^2}{p^2-m_6^2}\\
\Pi_{1t_R}&=\frac{y_R^2f^2}{2}\left(\frac{1}{p^2-m_6^2}-\frac{1}{p^2-m_4^2}\right)\\
\Pi_{1t_Lt_R}&=\frac{1}{2}y_Ly_Rf^2\left(-\frac{m_4}{p^2-m_4^2}+\frac{m_6}{p^2-m_6^2}\right)
\end{aligned}
\eea

\item $\textnormal{MCHM}_{14+14}$:\\
\bea
\begin{aligned}
\mathcal{L}_{14+14}&=y_L f (\bar{q}^{14}_L)^{ij}\left[U_{jJ}U_{iL}\Psi^{JL}_{9}+\sqrt{2}\ U_{i5}U_{jJ}\Psi^{J}_{4}+\frac{\sqrt{5}}{2}U_{i5}U_{j5}\Psi_1\right]\\
&\ \ +y_R f (\bar{t}^{14}_R)^{ij}\left[U_{jJ}U_{iL}\Psi^{JL}_{9}+\sqrt{2}\ U_{i5}U_{jJ}\Psi^{J}_{4}+\frac{\sqrt{5}}{2}U_{i5}U_{j5}\Psi_1\right]+\textnormal{h.c.}\\
&\ \ -m_{9}\bar{\Psi}_{9}\Psi_{9}-m_4\bar{\Psi}_{4}\Psi_{4}-m_1\bar{\Psi}_{1}\Psi_{1}
\end{aligned}
\eea 
\bea
\begin{aligned}
\Pi_{0t_L}&=1-\frac{f^2y_L^2}{p^2-m_4^2}\\
\Pi_{1t_L}&=\frac{5}{4}f^2y_L^2\left(-\frac{1}{p^2-m_1^2}+\frac{2}{p^2-m_4^2}-\frac{1}{p^2-m_9^2}\right)\\
\Pi_{2t_L}&=\frac{1}{4}y_L^2f^2\left(\frac{5}{p^2-m_1^2}-\frac{8}{p^2-m_4^2}+\frac{3}{p^2-m_9^2}\right)\\
\Pi_{0t_R}&=1-\frac{f^2y_R^2}{p^2-m_1^2}\\
\Pi_{1t_R}&=\frac{5}{2}y_R^2f^2\left(\frac{1}{p^2-m_1^2}-\frac{1}{p^2-m_4^2}\right)\\
\Pi_{2t_R}&=\frac{5}{16}y_R^2f^2\left(-\frac{5}{p^2-m_1^2}+\frac{8}{p^2-m_4^2}-\frac{3}{p^2-m_9^2}\right)\\
\Pi_{1t_Lt_R}&=\frac{\sqrt{5}}{2}y_Ly_Rf^2\left(-\frac{m_1}{p^2-m_1^2}+\frac{m_4}{p^2-m_4^2}\right)\\
\Pi_{2t_Lt_R}&=\frac{\sqrt{5}}{8}y_Ly_Rf^2\left(\frac{5m_1}{p^2-m_1^2}-\frac{8m_4}{p^2-m_4^2}+\frac{3m_9}{p^2-m_9^2}\right)
\end{aligned}
\eea

\item $\textnormal{MCHM}_{14+1}$:\\
\begin{align}
\mathcal{L}_{14+1}&=y_L f (\bar{q}^{14}_L)^{ij}\left[U_{jJ}U_{iL}\Psi^{JL}_{9}+\sqrt{2}\ U_{i5}U_{jJ}\Psi^{J}_{4}+\frac{\sqrt{5}}{2}U_{i5}U_{j5}\Psi_1\right]+y_R f \bar{t}_R\Psi_1+\textnormal{h.c.}\nn\\
&\ \ -m_{9}\bar{\Psi}_{9}\Psi_{9}-m_4\bar{\Psi}_{4}\Psi_{4}-m_1\bar{\Psi}_{1}\Psi_{1}
\end{align}
\bea
\begin{aligned}
\Pi_{0t_L}&=1-\frac{f^2y_L^2}{p^2-m_4^2}\\
\Pi_{1t_L}&=\frac{5}{4}f^2y_L^2\left(-\frac{1}{p^2-m_1^2}+\frac{2}{p^2-m_4^2}-\frac{1}{p^2-m_9^2}\right)\\
\Pi_{2t_L}&=\frac{1}{4}y_L^2f^2\left(\frac{5}{p^2-m_1^2}-\frac{8}{p^2-m_4^2}+\frac{3}{p^2-m_9^2}\right)\\
\Pi_{0t_R}&=1-\frac{f^2y_R^2}{p^2-m_1^2}\\
\Pi_{t_Lt_R}&=-\frac{\sqrt{5} m_1}{2}\cdot\frac{y_Ly_Rf^2}{p^2-m^2_1} s_hc_h
\end{aligned}
\eea

\item $\textnormal{CTHM}_{8+1}$:\\
\bea
\begin{aligned}
\mathcal{L}_{8+1}&=y_Lf (\bar{q}^8_L)^i \left[U_{iJ} \Psi_7^J+U_{i8}\Psi_1\right]+y_Rf\bar{t}_R\Psi_{1L}+\textnormal{h.c.}-m_7\bar{\Psi}_7\Psi_7-m_1\bar{\Psi}_1\Psi_1\\
&\ \ \ +\textnormal{twin\ sector}(y,m,\Psi\rightarrow \widetilde{y},\widetilde{m},\widetilde{\Psi})
\end{aligned}
\eea
\bea
\begin{aligned}
\Pi_{0t_L} &=1-\frac{y^2_Lf^2}{p^2-m_7^2}\\
\Pi_{1t_L} &=\frac{y_L^2f^2}{2}\left(\frac{1}{p^2-m^2_7}-\frac{1}{p^2-m^2_1}\right)\\
\Pi_{0t_R} &=1-\frac{y^2_Rf^2}{p^2-m_1^2}\\
\Pi_{1t_Lt_R} &=-\frac{m_1}{\sqrt{2}}\cdot\frac{y_Ly_Rf^2}{p^2-m^2_1}\\
\end{aligned}
\eea

\item $\textnormal{CTHM}_{8+28}$:\\
\bea
\begin{aligned}
\mathcal{L}_{8+28}&=y_Lf (\bar{q}^8_L)^i \left[U_{iJ} \Psi_7^J+U_{i8}\Psi_1\right]+y_R f (\bar{t}^{28}_R)^{ij}\left[U_{jJ}U_{iL}\Psi^{JL}_{21}+\sqrt{2} \ U_{i8}U_{jJ}\Psi^{J}_{7}\right]+\textnormal{h.c.}\\
&\ -m_{21}\bar{\Psi}_{21}\Psi_{21}-m_7\bar{\Psi}_{7}\Psi_{7}-m_1\bar{\Psi}_{1}\Psi_{1}+\textnormal{twin\ sector}(y,m,\Psi\rightarrow \widetilde{y},\widetilde{m},\widetilde{\Psi})
\end{aligned}
\eea  
\bea
\begin{aligned}
\Pi_{0t_L}&=1-\frac{y_L^2f^2}{p^2-m^2_7}\\
\Pi_{1t_L}&=\frac{y_L^2f^2}{2}\left(\frac{1}{p^2-m^2_7}-\frac{1}{p^2-m^2_1}\right)\\
\Pi_{0t_R}&=1-\frac{y_R^2f^2}{p^2-m^{2}_{21}}\\
\Pi_{1t_R}&=\frac{y_R^2f^2}{2}\left(\frac{1}{p^2-m^2_{21}}-\frac{1}{p^2-m^2_7}\right)\\
\Pi_{1t_Lt_R}&=-\frac{m_7}{2}\cdot\frac{y_Ly_Rf^2}{p^2-m^2_7}
\end{aligned}
\eea

\item $\textnormal{CTHM}_{8+35}$:\\
\bea
\begin{aligned}
\mathcal{L}_{8+35}&=y_Lf (\bar{q}^8_L)^i \left[U_{iJ} \Psi_7^J+U_{i8}\Psi_1\right]\\
&\ +y_R f (\bar{t}^{35}_R)^{ij}\left[U_{jJ}U_{iL}\Psi^{JL}_{27}+\sqrt{2}\ U_{i8}U_{jJ}\Psi^{J}_{7}+\sqrt{\frac{8}{7}}\ U_{i8}U_{j8}\Psi_1\right]+\textnormal{h.c.}\\
&\ -m_{27}\bar{\Psi}_{27}\Psi_{27}-m_7\bar{\Psi}_{7}\Psi_{7}-m_1\bar{\Psi}_{1}\Psi_{1}+\textnormal{twin\ sector}(y,m,\Psi\rightarrow \widetilde{y},\widetilde{m},\widetilde{\Psi})
\end{aligned}
\eea 
\bea
\begin{aligned}
\Pi_{0t_L}&=1-\frac{y_L^2f^2}{p^2-m^2_7}\\
\Pi_{1t_L}&=\frac{y_L^2f^2}{2}\left(\frac{1}{p^2-m^2_7}-\frac{1}{p^2-m^2_1}\right)\\
\Pi_{0t_R}&=1-\frac{y_R^2f^2}{7}\left(\frac{6}{p^2-m^2_{27}}+\frac{1}{p^2-m^2_1}\right)\\
\Pi_{1t_R}&=\frac{y_R^2f^2}{7}\left(\frac{3}{p^2-m^2_{27}}-\frac{7}{p^2-m^2_7}+\frac{4}{p^2-m^2_1}\right)\\
\Pi_{2t_R}&=\frac{y_R^2f^2}{7}\left(-\frac{3}{p^2-m^2_{27}}+\frac{7}{p^2-m^2_7}-\frac{4}{p^2-m^2_1}\right)\\ 
\Pi_{1t_Lt_R}&=\frac{y_Ly_Rf^2}{\sqrt{14}}\left(\frac{\sqrt{7}\ m_7}{p^2-m^2_7}-\frac{m_1}{p^2-m^2_1}\right)\\
\Pi_{2t_Lt_R}&=\frac{y_Ly_Rf^2}{\sqrt{14}}\left(\frac{2\ m_1}{p^2-m^2_1}-\frac{\sqrt{7}\ m_7}{p^2-m^2_7}\right)
\end{aligned}
\eea

\item $\textnormal{CMNNM}$:\\
\bea
\mathcal{L}&=&\ yf\bar{Q}_LU\Psi_R-M\bar{\Psi}_{L}\Psi_{R}-m\bar{\Psi}_{1L}t_R\\
&+&\widetilde{y}f\bar{\widetilde{Q}}_LU\widetilde{\Psi}_R-\widetilde{M}\bar{\widetilde{\Psi}}_{L}\widetilde{\Psi}_{R}-\widetilde{m}\bar{\widetilde{\Psi}}_{1L}\widetilde{T}_R-\widetilde{m}_{q}\bar{\widetilde{Q}}_{L}\widetilde{Q}_{R}+\text{h.c.}\ ,\nn
\eea
\begin{align}
&\Pi_{t_L}=1-\frac{y^2f^2}{2}\frac{1}{p^2-M^2}, \nn\\
& \Pi_{t_R}=1-\frac{m^2}{p^2-M^2},\nn\\
&\Pi_{t_Lt_R}=\frac{iyf}{\sqrt{2}}s_h\frac{Mm}{p^2-M^2}, \nn\\
&\widetilde{\Pi}_L=
\left(
\baa{cc}
1-\frac{\widetilde{y}^2f^2}{2(p^2-\widetilde{M}^2)}& 0\\ 
0 & 1-\frac{\widetilde{y}^2f^2}{p^2-\widetilde{M}^2}
\eaa
\right),\ \nn\\
&\widetilde{\Pi}_R=
\left(
\baa{cc}
1& 0\\ 
0 & 1-\frac{\widetilde{m}^2}{p^2-\widetilde{M}^2}
\eaa
\right),\ \nn\\
&\widetilde{\Pi}_{LR}=
\left(
\baa{cc}
\widetilde{m}_{q}& \frac{-i\widetilde{y}f}{\sqrt{2}}s_h\frac{\widetilde{m}\widetilde{M}}{p^2-\widetilde{M}^2}\\ 
0 & \widetilde{y}fc_h\frac{\widetilde{m}\widetilde{M}}{p^2-\widetilde{M}^2}
\eaa
\right).
\end{align}
\end{itemize}

\section{Higgs Couplings in Concrete Composite Models\label{sec:exp_coup}}
In this part, we collect the results of Higgs couplings in concrete composite Higgs models up to the leading order of $\mathcal{O}(\xi)$.
\begin{table}[!hbp]
\begin{tabular}{cc}
\hline
Couplings & Results\\
\cline{1-2}
$c_g$ & $1-\frac{1}{2}\xi$\\
\cline{1-2}
$c_{gghh}$ & $1$\\
\cline{1-2}
$c_t$ & $1-\frac{1}{2}\xi-\xi\left(\frac{\frac{1}{2}y_L^2f^2(m_4^2-m_1^2)}{m_1^2m_4^2+m_1^2y_L^2f^2}\right)$\\
\cline{1-2}
$c_{t\bar{t}hh}$ & $ -\frac{\xi}{2}-\frac{3}{2}\xi \left(\frac{\frac{1}{2}y_L^2f^2(m_4^2-m_1^2)}{m_1^2m_4^2+m_1^2y_L^2f^2}\right)$\\
\cline{1-2}
$m_t$ & $ \sqrt{\xi} \frac{\frac{1}{\sqrt{2}}y_Ly_Rf^2m_4}{\sqrt{(m_4^2+y_L^2f^2)(m_1^2+y_R^2f^2)}} $\\
\hline
\end{tabular}
\caption{Higgs Couplings and top mass in MCHM of $5+1$ Representation, which means $t_L$ is embedded in the $5$ of $SO(5)$ while $t_R$ is a singlet.}
\end{table}

\begin{table}[!hbp]
\begin{tabular}{cc}
\hline
Couplings & Results\\
\cline{1-2}
$c_g$ & $1-\frac{3}{2} \xi$\\
\cline{1-2}
$c_{gghh}$ & $1+\xi$\\
\cline{1-2}
$c_t$ & $1-\frac{3}{2}\xi-\xi \left(\frac{\frac{1}{2}y_L^2f^2\left(m_4^2-m_1^2\right)}{m^2_1m^2_4+m^2_1y_L^2f^2}+\frac{y_R^2f^2\left(m_1^2-m_4^2\right)}{m^2_4m^2_1+m_4^2y_R^2f^2}\right)<1-\frac{3}{2}\xi$\\
\cline{1-2}
$c_{t\bar{t}hh}$ & $-2\xi-\frac{3}{2}\xi \left(\frac{\frac{1}{2}y_L^2f^2\left(m_4^2-m_1^2\right)}{m^2_1m^2_4+m^2_1y_L^2f^2}+\frac{y_R^2f^2\left(m_1^2-m_4^2\right)}{m^2_4m^2_1+m_4^2y_R^2f^2}\right)<-2\xi$\\
\cline{1-2}
$m_t$ & $ \sqrt{\xi} \frac{\frac{1}{\sqrt{2}}y_Ly_Rf^2\bold\|-m_1+m_4\bold\|}{\sqrt{(m_4^2+y_L^2f^2)(m_1^2+y_R^2f^2)}}$\\
\hline
\end{tabular}
\caption{Higgs Couplings and top mass in MCHM of $5+5$ Representation, which means $t_L$ and $t_R$ are both embedded in the $5$ of $SO(5)$.}
\end{table}

\begin{table}[!hbp]
\begin{tabular}{cc}
\hline
Couplings & Results\\
\cline{1-2}
$c_g$ & $1-\frac{3}{2} \xi$\\
\cline{1-2}
$c_{gghh}$ & $1+\xi$\\
\cline{1-2}
$c_t$ & $1-\frac{3}{2}\xi-\xi \left(\frac{\frac{1}{2}y_L^2f^2\left(m_4^2-m_6^2\right)}{m^2_6m^2_4+m^2_6y_L^2f^2}+\frac{\frac{1}{2}y_R^2f^2\left(m_6^2-m_4^2\right)}{m^2_4m^2_6+m_4^2y_R^2f^2}\right)<1-\frac{3}{2}\xi$\\
\cline{1-2}
$c_{t\bar{t}hh}$ & $-2\xi-\frac{3}{2}\xi \left(\frac{\frac{1}{2}y_L^2f^2\left(m_4^2-m_6^2\right)}{m^2_6m^2_4+m^2_6y_L^2f^2}+\frac{\frac{1}{2}y_R^2f^2\left(m_6^2-m_4^2\right)}{m^2_4m^2_6+m_4^2y_R^2f^2}\right)<-2\xi$\\
\cline{1-2}
$m_t$ & $ \sqrt{\xi} \frac{\frac{1}{2}y_Ly_Rf^2\bold\|-m_4+m_6\bold\|}{\sqrt{(m_4^2+y_L^2f^2)(m_6^2+y_R^2f^2)}}$\\
\hline
\end{tabular}
\caption{Higgs Couplings and top mass in MCHM of $10+10$ Representation, which means $t_L$ and $t_R$ are both embedded in the $10$ of $SO(5)$.}
\end{table}

\begin{table}[!hbp]
\begin{tabular}{cc}
\hline
Couplings & Results\\
\cline{1-2}
$c_g$ & $ 1+\xi \frac{3m_1m_4-11m_1m_9+8m_4m_9}{2m_1m_9-2m_4m_9}$\\
\cline{1-2}
$c_{gghh}$ & $1-\xi \frac{3m_1m_4-10m_1m_9+7m_4m_9}{2m_1m_9-2m_4m_9}$\\
\cline{1-2}
$c_t$ & $1+\frac{\xi}{4}\left(-6-\frac{5f^2y_L^2(m_1^2m_4^2-2m_1^2m_9^2+m_4^2m_9^2)}{m_1^2m_9^2(f^2y_L^2+m_4^2)}+\frac{10f^2y_R^2(m_4^2-m_1^2)}{m_4^2(f^2y_R^2+m_1^2)}+\frac{6m_1m_4-16m_1m_9+10m_4m_9}{m_1m_9-m_4m_9}\right) $\\
\cline{1-2}
$c_{t\bar{t}hh}$ & $ -2\xi+\frac{9m_1m_4-24m_1m_9+15m_4m_9}{4m_1m_9-4m_4m_9}\xi-\frac{15\xi}{8}\left(\frac{y_L^2f^2(m_1^2m_4^2-2m_1^2m_9^2+m_4^2m_9^2)}{m_9^2m_1^2(f^2y_L^2+m_4^2)}+\frac{2y_R^2f^2(m_1^2-m_4^2)}{m_4^2(m_1^2+f^2y_R^2)}\right)$\\
\cline{1-2}
$m_t$ & $ \sqrt{\xi}  \frac{\frac{\sqrt{5}}{2}f^2y_Ly_R \bold\|m_4-m_1\bold\|}{\sqrt{(m_1^2+f^2y_R^2)(m_4^2+f^2y_L^2)}}$\\
\hline
\end{tabular}
\caption{Higgs Couplings and top mass in MCHM of $14+14$ Representation, which means $t_L$ and $t_R$ are both embedded in the $14$ of $SO(5)$.}
\end{table}

\begin{table}[!hbp]
\begin{tabular}{cc}
\hline
Couplings & Results\\
\cline{1-2}
$c_g$ & $1-\frac{3}{2}\xi$\\
\cline{1-2}
$c_{gghh}$ & $1+\xi$\\
\cline{1-2}
$c_t$ & $1-\frac{3}{2}\xi-\xi\frac{5f^2y_L^2}{4}\frac{\left(\frac{1}{m_1^2}-\frac{2}{m_4^2}+\frac{1}{m_9^2}\right)}{\left(1+\frac{f^2y_L^2}{m_4^2}\right)}$\\
\cline{1-2}
$c_{t\bar{t}hh}$ & $-2\xi-\xi \frac{15f^2y_L^2\left(\frac{1}{m_1^2}-\frac{2}{m_4^2}+\frac{1}{m_9^2}\right)}{8\left(1+\frac{f^2y_L^2}{m_4^2}\right)}$\\
\cline{1-2}
$m_t$ & $ \frac{\sqrt{5\xi}f^2 y_L y_R}{2 m_1 \sqrt{\left(1+\frac{y_R^2f^2}{m_1^2}\right)\left(1+\frac{y_L^2f^2}{m_4^2}\right)}} $\\
\hline
\end{tabular}
\caption{Higgs Couplings and top mass in MCHM of $14+1$ Representation, which means $t_L$ is embedded in the $14$ of $SO(5)$ while $t_R$ is a singlet.}
\end{table}

\begin{table}[!hbp]
\begin{tabular}{cc}
\hline
Couplings & Results\\
\cline{1-2}
$c_g$ & $1-\frac{1}{2}\xi$\\
\cline{1-2}
$c_{gghh}$ & $1$\\
\cline{1-2}
$c_t$ & $1-\frac{1}{2}\xi-\xi\left(\frac{\frac{1}{2}y_L^2f^2(m_7^2-m_1^2)}{m_1^2m_7^2+m_1^2y_L^2f^2}\right)$\\
\cline{1-2}
$c_{t\bar{t}hh}$ & $ -\frac{\xi}{2}-\frac{3}{2}\xi \left(\frac{\frac{1}{2}y_L^2f^2(m_7^2-m_1^2)}{m_1^2m_7^2+m_1^2y_L^2f^2}\right)$\\
\cline{1-2}
$m_t$ & $ \sqrt{\xi} \frac{\frac{1}{\sqrt{2}}y_Ly_Rf^2m_7}{\sqrt{(m_7^2+y_L^2f^2)(m_1^2+y_R^2f^2)}} $\\
\hline
\end{tabular}
\caption{Higgs Couplings and top mass in CTHM of $8+1$ Representation, which means $t_L$ is embedded in the $8$ of $SO(8)$ while $t_R$ is a singlet. MCHM of $5+1$ representation is very similar to this case.}
\end{table}

\begin{table}[!hbp]
\begin{tabular}{cc}
\hline
Couplings & Results\\
\cline{1-2}
$c_g$ & $1-\frac{1}{2}\xi$\\
\cline{1-2}
$c_{gghh}$ & $1$\\
\cline{1-2}
$c_t$ & $1-\frac{1}{2}\xi-\xi\left(\frac{\frac{1}{2}y_L^2f^2(m_7^2-m_1^2)}{m_1^2m_7^2+m_1^2y_L^2f^2}+\frac{\frac{1}{2}y_R^2f^2(m_{21}^2-m_7^2)}{m_7^2m_{21}^2+m_7^2y_R^2f^2}\right)$\\
\cline{1-2}
$c_{t\bar{t}hh}$ & $ -\frac{\xi}{2}-\frac{3}{2}\xi \left(\frac{\frac{1}{2}y_L^2f^2(m_7^2-m_1^2)}{m_1^2m_7^2+m_1^2y_L^2f^2}+\frac{\frac{1}{2}y_R^2f^2(m_{21}^2-m_7^2)}{m_7^2m_{21}^2+m_7^2y_R^2f^2}\right)$\\
\cline{1-2}
$m_t$ & $ \sqrt{\xi} \frac{\frac{1}{2}y_Ly_Rf^2m_{21}}{\sqrt{(m_7^2+y_L^2f^2)(m_{21}^2+y_R^2f^2)}}$\\
\hline
\end{tabular}
\caption{Higgs Couplings and top mass in CTHM of $8+28$ Representation, which means $t_L$ is embedded in the $8$ of $SO(8)$ while $t_R$ is embedded in the $28$ of $SO(8)$.}
\end{table}

\begin{table}[!hbp]
\begin{tabular}{cc}
\hline
Couplings & Results\\
\cline{1-2}
$c_g$ & $1-\frac{1}{2}\xi+2\xi\frac{\sqrt{7}\ m_1-2\ m_7}{m_7-\sqrt{7}\ m_1}$\\
\cline{1-2}
$c_{gghh}$ & $1-2\xi\frac{\sqrt{7}\ m_1-2\ m_7}{m_7-\sqrt{7}\ m_1}$\\
\cline{1-2}
$c_t$ & $1-\frac{1}{2}\xi-\frac{\xi}{14}f^2\left(\frac{2y_R^2\left(-\frac{4}{m_1^2}-\frac{3}{m_{27}^2}+\frac{7}{m_7^2}\right)}{1-\frac{1}{7}f^2y_R^2\left(-\frac{1}{m_1^2}-\frac{6}{m_{27}^2}\right)}+\frac{7y_L^2\left(\frac{1}{m_1^2}-\frac{1}{m_7^2}\right)}{1-\frac{f^2y_L^2}{m_7^2}}\right)+2\xi\frac{\sqrt{7}\ m_1-2\ m_7}{m_7-\sqrt{7}\ m_1}$\\
\cline{1-2}
$c_{t\bar{t}hh}$ & $ -\frac{\xi}{2}+3\xi\frac{\sqrt{7}\ m_1-2\ m_7}{m_7-\sqrt{7}\ m_1}-\frac{3}{2}\frac{\xi}{14}f^2\left(\frac{2y_R^2\left(-\frac{4}{m_1^2}-\frac{3}{m_{27}^2}+\frac{7}{m_7^2}\right)}{1-\frac{1}{7}f^2y_R^2\left(-\frac{1}{m_1^2}-\frac{6}{m_{27}^2}\right)}+\frac{7y_L^2\left(\frac{1}{m_1^2}-\frac{1}{m_7^2}\right)}{1-\frac{f^2y_L^2}{m_7^2}}\right)$\\
\cline{1-2}
$m_t$ & $\frac{f^2\sqrt{\xi}y_Ly_R\left(\frac{1}{m_1}-\frac{\sqrt{7}}{m_7}\right)}{\sqrt{14}\sqrt{\left(1-\frac{y_L^2f^2}{m_7^2}\right)\left(1-\frac{y_R^2f^2}{7}\left(-\frac{1}{m_1^2}-\frac{6}{m_{27}^2}\right)\right)}}$\\
\hline
\end{tabular}
\caption{Higgs Couplings and top mass in CTHM of $8+35$ Representation, which means $t_L$ is embedded in the $8$ of $SO(8)$ while $t_R$ is embedded in the $35$ of $SO(8)$.}
\end{table}

\begin{table}[!hbp]
\begin{tabular}{cc}
\hline
Couplings & Results\\
\cline{1-2}
$c_g$ & $1-\frac{1}{2}\xi$\\
\cline{1-2}
$c_{gghh}$ & $1$\\
\cline{1-2}
$c_t$ & $1-\frac{1}{2}\xi$\\
\cline{1-2}
$c_{t\bar{t}hh}$ & $ -\frac{\xi}{2}$\\
\cline{1-2}
$m_t$ & $ \sqrt{\xi} \frac{\frac{1}{\sqrt{2}}y_LfmM}{\sqrt{(M^2+y_L^2f^2)(M^2+m^2)}} $\\
\hline
\end{tabular}
\caption{Higgs Couplings and top mass in CMNNM. MCHM of $5+1$ representation and CTHM of $8+1$ representation are very similar to this case.}
\end{table}

\clearpage
\bibliography{compositehiggs}
\end{document}